\documentstyle[12pt]{article}
\begin{document}
\input epsf.sty
\begin{titlepage}
\title{ \bf Quantum description of  the orientational degrees of  
freedom in a
biaxial nematic liquid}
\author{ Jorge
Alfaro \thanks{e-mail
address:
jalfaro@lascar.puc.cl .} , Oscar Cubero
\\ Facultad de F\'\i sica\\Universidad Cat\'olica de Chile\\Casilla 306,
Santiago 22, Chile\\
{} \\
and \\ {} \\ Luis F.  Urrutia  \\
Departamento de F\'\i sica de Altas Energ\'\i as \\ Instituto de Ciencias
Nucleares\\
Universidad Nacional
Aut\'onoma de M\'exico \\ Apartado Postal
70-543,
04510
M\'exico, D.F.  }
\maketitle

\begin{abstract}
The quantum mechanical version of a classical model for
studying the orientational
degrees of freedom corresponding to a  nematic liquid composed of biaxial
molecules  is presented. The effective degrees of freedom  are   
described by
operators carrying an $SU(3)$ representation, which allows the explicit
calculation of the partition function in the mean field  
approximation. The
algebraic consistency conditions are solved  numerically and the  
equilibrium
phases of the system are determined. In particular, the entropy,   
the specific
heat and the order parameters
are presented for different choices of the constituent biaxial molecules. Our results reproduce the
classical calculation
in the limit of high temperatures and high quantum numbers.
\end{abstract}
%\vfill
%\vspace{11cm}
\begin{flushleft}
%IFT-P.014/97\\

\end{flushleft}
\end{titlepage}

\newpage
\section{Introduction}

Nematic liquids constitute  a particular case of the phases  
denominated with
the generic name of liquid crystals \cite{REV,CHAILUB}. This is a  
state of
matter which is intermediate between the liquid and the solid  
phases. It lacks
the positional symmetry characterizing  the lattice of a crystal,  
which makes
it similar to a liquid, but it retains some orientational symmetry  
that is not
present in a normal (isotropic) liquid. The molecules forming a  
liquid crystal
can be roughly caracterized as
rod-like or disc-like, and the corresponding orientational symmetry
is manifested by different types of local aligments of such  
molecules with
respect to a preferred direction characterized by a unit vector  
called the
director. In the case of a nematic liquid, the director is
fixed in space and the molecules remain, on the average, parallel  
to it, though
they can difuse as in a normal liquid. The existence of
a preferred direction makes this phase to exhibit birrefringent  
properties. On
the other hand, the energies
involved in the transition from the liquid crystal phase to the
isotropic liquid phase are normally small compared with the  
corresponding ones
in the solid-liquid transition. This makes it possible to  
drastically change
the response of liquid crystals by external agents, using  
relatively small
amounts of energy. It is this property which determines the numerous and
important technological
applications of liquid crystals \cite{COLLINS}.

Besides the important applications of liquid crystals in everyday's life,
together with the huge amount of theoretical work that has been  
consequently
developped, it has been suggested  recently that these phases might  
be relevant
also  in other  situations, like for example, the description of  nuclear
matter under very stringent conditions of densities and temperatures
\cite{MIGDAL},  particularly in the case of neutron stars.  These
quantum objects  
are cold
stars composed primarily of neutrons and supported against collapse  
by neutron
degeneracy pressure.  In this situation,
it is predicted  that the  inner crust of a neutron star (roughly
located at $9.2$ km from the center of a typical neutron star with  
a radious of
$10$ km and a mass of $1.5$ solar masses) is separated from the  
core, composed
mainly of uniform nuclear matter,  by  a thin spherical shell about  
$0.1$ km
wide, containing nuclear matter in the form of non-spherical nuclei  
(rods,
plates, tubes) plus dripped neutrons plus electrons. This  liquid crystal
behavior, as opposed to a  crystalline solid, will  have important  
consequences
for a number of aspects of  neutron star properties \cite{PETHICK}.
Nevertheless, there is  a  main difference between this case and  
the situation
in the laboratory: in the neutron star case, the basic entities  
from which
nuclei  are constructed are nucleons, which are spherical; while
in laboratory liquid crystals, the basic ingredientes are  
non-spherical biaxial
molecules in general \cite{PETHICK}.  In this  respect, it might be  
interesting to consider  liquid crystals made up from  constituentes having   
permanent
magnetic moments, as nucleons,  generating a liquid ferromagnet. In  the
laboratory, such structures, though not forbidden,  should be very  
rare because
the exchange couplings beteween  neighboring magnetic moments are  
much weaker
than the conventional interactions between molecules.

%%%%%%%%%%%%%%%%%%%%%%%%%%%%%%%%%%%%%%%%%%%%%%%

Motivated by  some of the above considerations, we consider the quantum
mechanical  description of  the orientational degrees of freedom of  
a nematic
liquid
consisting of biaxial molecules, in the  mean field  approximation.  
This system
has been previously
discussed in Ref.\cite{MULDER}, where an exact calculation of the  
classical
partition function, together with an extensive analysis
of the corresponding phases, transition temperatures,
themodynamic quantities and
order parameters, was made. We will refer to this work as the
classical model.  In this reference, the effective  
orientational
degrees of
freedom of the non-spherically symmetric molecules were described
by an $SU(3)$
invariant
Hamiltonian constructed  from the angular momentum operators
$L_i, \ i=1,2,3$ together with the traceless
quadrupole tensor operator
$Q_{\alpha \beta}, \quad \alpha, \beta=1,2,3$, which constitute a
representation
for the generators of $SU(3)$. The interpretation of the model  was  
performed
in a
coordinate
system where the molecule is at rest, i.e. $L_i=0$. In this system, the
quadrupole operator ${\bar Q}_{\alpha \beta}$ relates to the
inertia tensor of
the molecule in such
a way that the $SU(3)$ Casimir operators, related to the eigenvalues
of ${\bar Q}_{\alpha \beta}$, provide a description of the
shape of the molecule.

The classical partition function in the mean field
approximation was exactly calculated in Ref.\cite{MULDER} by
first rewritting it
as an  integral
over the space parameters of the $SU(3)$ group, according to a
theorem due to Macfarlane et al. \cite{MAC}. This integral
has been  previously calculated in
the bosonic case  by Harish-Chandra, and  Itzykson
and Zuber \cite{IZ}.  The exact evaluation of the partition  
function,
together
with the fact that the consistency conditions  consisted in two  
 coupled
algebraic
equations, constituted an advantage  over previous statistical   theories
proposed  \cite{PT} in order to explain the  isotropic  
$\rightarrow$ uniaxial
$\rightarrow$ biaxial phase transitions discovered experimentally  
in nematic
liquids \cite{EXP}. The works included in \cite{PT}
are  generalizations of the Maier-Saupe theory, that  was for some  
time the
fullest existing statistical mechanical  theory of the nematic  
state, which,
nevertheless,  did not include the biaxial phase
\cite{MS}.

Following an approach similar to some previous works, we only  
consider the
orientational interaction energy among the molecules of the liquid,  
 in such a
way that  the exact nature  of the intermolecular forces need not   
be specified
. Whether the liquid  crystal state arises because of the hard core  
repulsive
forces or the long range dispersive forces is a matter of  
indifference in the
following. This approximation is expected to produce  qualitatively  
reasonable
results in the nematic case \cite{PT}. In this way, we are  really  
calculating
the excess thermodynamical properties of the ordered system,  
relative to those
of the  disordered one.

In this work we present a  quantum description of the problem  
discussed in
\cite{MULDER},
which amounts
to the generalization of the  usual Heisenberg model  describing  the
interaction of
angular momentum  operators ($SU(2)$ case ) to the situation where the
interacting operators carry an $SU(3)$
representation.  Similarly to the classical case, we are able to exactly
calculate the partition function in the mean field approximation. This is
achieved by using the  Weyl formula for the characters of  $SU(3)$  
\cite{WEYL}.
  We expect to recover the results of
Ref.\cite{MULDER}  in the
high temperature, high quantum numbers limit  \cite{LIEB,SIMON}.

The paper is organized as follows. In section 2 we introduce the  
model and
apply the Weyl formula to calculate the partition function.  Section 3
describes the characterization of the constituents molecules of the  
liquid in
terms of the irreducible representations of $SU(3)$. The equilibrium  
phase is
parametrized in the same way as in  Ref. \cite{MULDER} as briefly  
recalled in
section 4.  The connection with the classical case is presented in  
section 5,
where a function relating the corresponding partitions functions is  
explicitly
calculated. Section 6  contains the
selfconsistent equations that allow for a numerical determination of the
equilibrium parameters,
which are obtained by determining the absolute mimima of the free  
energy of the
system. Besides, some symmetry properties of the quantum partition  
function are
exhibited there. Section 7  describes the calculation of the equilibrium
parameters of the different phases of the system, emphasizing those  
limiting
cases that can be treated algebraically. We have also calculated  
numerically
the specific heat and the entropy as functions of the adimensional  
temperature,
for
some representative cases. The order parameters in the quantum case  
are also
defined in this section.  In many of  
the figures
presented, we have shown
both the classical and quantum predictions in order to assess the main
differences between the two
approaches.  Some  conclusions and  comments are presented in section 9.  
Finally, there is
an Appendix summarizing some basic properties of $SU(3)$. Here we  
also give
an alternative way of rewritting  the  partition function, in  terms of
polynomials, which have been useful in obtaining  
some limiting
cases considered  along  this work.

\section{ The model}

Following the work of Ref. \cite{MULDER}, let us consider the effective
Hamiltonian
\begin{equation}
\label{HAM}
 H=-\frac{J}{N}\sum_{m\neq n}S_{a}^{m}S_{a}^{n},
\end{equation}
which describes the orientational energy of the system. Here,  the $N$
independent operators $ S^m_a, \ m=1,2,3, \dots, N,
\ a =1,2,3,
\dots, 8$ describe the degrees of freedom of each molecule of the
liquid
and  carry an $SU(3)$ representation, i.e.
\begin{equation}\label{ALG}
[ S^m_a, \ S^n_b]= \delta^{m n} i f_{abc} \  S^m_c,
\end{equation}
where $f_{abc}$ are the $SU(3)$ structure constants. As it is well
known, the
fundamental representation of this group is written in terms of
the Gell-Mann
$3 \times 3$ matrices
$\lambda_a$  given in the Appendix, where other useful properties
are also
recalled.

In the mean field (molecular field) aproximation we replace
(\ref{HAM}) by
\begin{equation}
\label{HAMMF}
 H=-{J} B_a   \sum_{n}S_{a}^{n},
\end{equation}
where the eight-dimensional vector  $B_a=\frac{1}{N}< \sum_{m}S_a^m>$,
which represents the average field seen by each molecule,  has to
be determined
consistently.

In this approximation, the partition function $Z$ is
\begin{equation}\label{Z}
Z=Tr e^ {-\beta H}=Tr   e^ {\beta
{J}B_{a}  \sum_{n=1}^{N}S_{a}^{n}}=
Z_{0}^N,
\end{equation}
where
\begin{equation}\label{Z0}
 Z_{0}=Tr e^ {\beta
{J} B_{a}  S_{a}}.
\end{equation}
As usual  $\frac{1}{\beta} ={\rm k}{\cal T}$,  where ${\rm k}$ is  
the Boltzman
constant and ${\cal T}$
is the temperature of the system.

The consistency equations for  the mean field approximation reduce to
\begin{equation}
\label{CONSEQ}
B_a=\frac{1}{\beta J}\frac{\partial}{\partial B_a} ln Z_0.
\end{equation}

The basic  problem now  is to calculate such partition function for an
arbitrary
$(p,q)$ irreducible representation of $SU(3)$. The first  step is
to select and
adequate coordinate system to perform the calculation. Since the trace is
invariant under a unitary transformation
of the operator inside, we can diagonalize  the hermitian operator
$ {\beta
{J} B_{a}  S_{a}}$ in such a way that  it can be written as $ \beta
J (B_3 S_3
+ B_8 S_8)$, where we  assume that the operators $S_3$ and
$S_8$ are
chosen to be simultaneously diagonal. This is equivalent to choose
a coordinate
system where the eight-dimensional
vector $B_a$ has only  non-zero components  in the $3$ and $8$ axis.
The fact
that we need
two non-zero components, instead of one as in the $SU(2)$ case, is
related to
the property that in
the adjoint representation of $SU(3)$ there are two  invariants:  $
B_aB_a$
and
$d_{a b c}B_aB_bB_c$. Using the explicit form of $d_{a b c}$, it
can be shown
that it is possible
to select the $B_3$ and $B_8$ components as the only non-zero ones. In
this way we are left with
\begin{equation}
\label{Z0RED}
Z_0=Tr e^{\beta J (B_3 S_3 + B_8 S_8)}.
\end{equation}
 From now on we will assume that $B_a= ( 0,0,B_3,0, \dots, B_8)$.

In the fundamental representation $(1,0)$, the partition function
is given by
\begin{equation}
\label{ZFUND}
Z_{0}=\sum_{{i}}e^{\beta J {\vec B} \cdot{\vec \mu}_{i}},
\end{equation}
where ${\vec \mu}_i$ are the weights of the fundamental representation.

The partition function (\ref{Z0}) is the character
of the operator $e^{-\beta H}$. In this way, to calculate $Z_0$  for
an arbitrary representation  $(p,q)$ of  $SU(3)$,  we can use
the Weyl formula for  the characters
of $SU(n)$, given in terms of the eigenvalues of the
operator in the fundamental representation. For $SU(3)$, the Weyl
formula reduces to \cite{WEYL}
\begin{equation}
\label{WEYL}
Z_{0}\equiv\Sigma^{\{\eta_{1},\eta_{2}\}}=
\frac{1}{\left(\Lambda_{1}-\Lambda_{2}\right)
      \left(\Lambda_{1}-\Lambda_{3}\right)
      \left(\Lambda_{2}-\Lambda_{3}\right)}
\left|\begin{array}{ccc}
   \Lambda_{1}^{\eta_{1}+2} & \Lambda_{1}^{\eta_{2}+1} & 1  \\
   \Lambda_{2}^{\eta_{1}+2} & \Lambda_{2}^{\eta_{2}+1} & 1  \\
   \Lambda_{3}^{\eta_{1}+2} & \Lambda_{3}^{\eta_{2}+1} & 1
\end{array} \right|.
\end{equation}

Here $ \eta_{1}=p+q $ is the number of boxes in the first row
of the corresponding Young tableau, while $ \eta_{2}=q $
is the number of boxes in the second row of the tableau. In other words,
$p$ is the number of columns with one box $( p \geq 0)$  and $q$
is the number of columns with two boxes $(q \geq 0)$ in the tableau. The
representation $(0,0)$ is the trivial one. The
eigenvalues
$\Lambda_i, \ i=1,2,3$ \ are given by
\begin{equation}
\label{EIGEN}
\Lambda_i= e^{\beta J {\vec B} \cdot {\vec \mu}_i}
\end{equation}
and satisfy the condition $\Lambda_1 \Lambda_2 \Lambda_3=1$. The
weights ${\vec \mu}_i$ are given in the  Appendix.

A more convenient notation is
\begin{equation}
\label{NOTAC}
\rho=\frac{1}{2}\beta J B_3,
\quad
\sigma=\frac{1}{2\sqrt{3}}\beta J B_8,
\end{equation}
which produces the following expressions for the eigenvalues of the
Hamiltonian in  the fundamental representation
\begin{equation}
\label{EIRED}
\Lambda_1= e^{\rho+ \sigma}, \quad \Lambda_2=e^{-\rho + \sigma},
\quad
{\Lambda_3}^{-1}=\Lambda_1 \Lambda_2= e^{2 \sigma}.
\end{equation}

The partition function is given by

\begin{eqnarray}
\label{Z0EXPL}
Z_{0}^{(p,q)}&=& \frac{1}{\sinh 2 \rho-2\sinh{\rho}
\cosh {3 \sigma}} \left(
- e^{(p+2q+3)\sigma}\ \sinh {(p+1)\rho}
 \right.
\nonumber \\
&& \left. +  e^{(p-q) \sigma} \ \sinh(p+q+2)\rho
- e^{ -(q+2p+3) \sigma} \ \sinh{(q+1)\rho}\right).
\nonumber \\
\end{eqnarray}
It is convenient to observe that the above expression for
$Z_0^{(p,q)}$ can be
rewritten as
\begin{equation}
\label{ZFRACC}
Z_0^{(p,q)}=\frac{N^{(p,q)}}{N^{(0,0)}},
\end{equation}
where
\begin{eqnarray}
\label{NUM}
N^{(p,q)}&=&e^{(p-q) \sigma} \ \sinh(p+q+2)\rho-
e^{(p+2q+3)\sigma}\ \sinh
{(p+1)\rho}\nonumber \\
&-&e^{-(q+2p+3) \sigma} \ \sinh{(q+1)\rho}.
\end{eqnarray}

\section{Specification of the constituent molecules}

The orientational degrees of freedom of the molecules are described by
the eight operators $S_a$, which can be reinterpreted as
the angular momentum $L_i, \ i=1,2,3$
and the quadrupole moment $Q_{ij}$ operators,
according to \cite{MULDER},
\begin{eqnarray}\label{RELDEG}
&&L_1= 2 S_7, \quad L_2=-2 S_5\quad L_3= 2 S_2, \nonumber \\
&& Q_{12}=S_1, \quad Q_{13}= S_4, \quad Q_{23}=S_6, \nonumber \\
&& Q_{11}= S_3+ \frac{1}{\sqrt{3}}S_8, \quad
Q_{22}=-S_3+ \frac{1}{\sqrt{3}}S_8, \quad  Q_{33}= -\frac{2}{\sqrt{3}}
S_8. \nonumber \\
\end{eqnarray}
The molecules forming the liquid will be characterized
by the representation $(p,q)$ of $SU(3)$. To this end
we consider the two Casimir operators of the group
\begin{equation}\label{CASIMIR}
{\bf C_{2}}=S_{a}S_{a}=I_{2}{\bf {I}} , \qquad
{\bf C_{3}}=d_{abc}S_{a}S_{b}S_{c}=I_{3}{\bf {I}}
\end{equation}
which take the following values in each irreducible representation
$ (p,q ) $, \cite{SHARP}

\begin{equation}\label{CASREP1}
 I_{2}=\frac{p^{2}+q^{2}+pq+3\left(p+q\right)}{3},
\end{equation}
\begin{equation}\label{CASREP2}
I_{3}=\frac{1}{18}(p-q)(p +2q+3)(2p+q+3).
\end{equation}

Motivated by the classical description and in order to compare with
it, we
introduce
a parametrization of the above invariants in terms of $\mu_i, \
i=1,2,3 $,  with
the restriction
$\mu_1 + \mu_2 + \mu_3=0$. In the classical case one defines
$I_3=\frac{3}{2}
\mu_1\mu_2\mu_3$, which can be reproduced in our case by choosing
\begin{equation}\label{QUANTEIG}
\mu_1= -\frac{1}{3}(p+ 2q + 3), \quad \mu_2= - \frac{1}{3}( p-q),
\quad \mu_3=
\frac{1}{3}(2p +q +3),
\end{equation}
according to the relation (\ref{CASREP2}). Following the analogy, we next
calculate what used to be the quadratic Casimir operator. We obtain
\begin{equation}\label{INVEIG}
-(\mu_1 \mu_2+ \mu_1 \mu_3+ \mu_2 \mu_3 )=I_2+1,
\end{equation}
where $I_2$ is given by Eq.(\ref{CASREP1}).

 The choice (\ref{QUANTEIG}) implies the ordering
$\mu_1 \leq \mu_2 \leq \mu_3$, for arbitrary values of $p$ and $q$.

The eigenvalues (\ref{QUANTEIG}) satisfy the cubic equation
\begin{equation}\label{EQEIG}
\mu^3 - (I_2 +1) \mu - \frac{2}{3}I_3=0,
\end{equation}
and can be parametrized as
\begin{equation}\label{EIGPAR}
\mu_1=\mu \ { \rm cos}\left( \psi +\frac{2\pi}{3} \right), \quad
\mu_2=\mu \ { \rm cos}\left( \psi -\frac{2\pi}{3} \right), \quad
\mu_3=\mu \ { \rm cos} \psi ,
\end{equation}
which is consistent with the ordering introduced in (\ref{QUANTEIG}).
To this end we must always have
\begin{equation}
\label{PROPAR}
\mu \ {\rm sin} \psi >0,\quad
\mu  \ {\rm cos} \psi >0, \quad {\rm tan}\psi \leq \sqrt{3},
\end{equation}
 where
\begin{equation}\label{PARCAS}
\mu= \pm 2 \sqrt{\frac{(I_2 +1)}{3}}, \quad {\rm cos} 3 \psi= \pm
\frac{\sqrt{3} I_3}{
{(I_2+ 1)}^{\frac{3}{2}}},
\end{equation}
in complete analogy with the classical case.
The above conditions lead to the following correlations
between $\psi$ and $\mu$
\begin{equation}
\label{PSIMU}
\mu >0 \leftrightarrow  \ 0 < \psi< \frac{\pi}{3}, \qquad
\mu <0 \leftrightarrow \ \pi < \psi< \pi +\frac{\pi}{3},
\end{equation}
 for arbitrary $p$ and $q$. In this way, for $p$ and $q$ fixed,
the change $\mu \rightarrow - \mu$ has to be followed by $\psi
\rightarrow \psi + \pi$. Thus we pass from the upper
plus sign to the lower minus sign in the two relations contained in
(\ref{PARCAS}).

The  physical interpretation associated to the $(p,q)$
characterization of the
individual molecules is made through the classical connection

\begin{equation}\label{NUSHAPE}
a_i{}^2= \frac{1}{3}( a_1{}^2 + a_2{}^2 +a_3{}^2) + \frac{1}{2}\mu_i,
\end{equation}
which relates the  $\mu$-values (\ref{QUANTEIG}) with the shape
corresponding
to a molecule described by a mass distribution defined by six equal
masses at
distances $ \pm a_1,  \pm a_2,  \pm a_3 $ from the center  
\cite{MULDER}. Here
we are assuming that the ${a_i}$'s are
written in dimensionless units.  The
correspondence  is
\begin{eqnarray}\label{PHASES}
&&\mu_1= \mu_2=\mu_3  \quad   ({\rm  spherically \ \ symmetric \ \
molecule}),
\nonumber \\
&& \mu_1= \mu_2  < 0, \quad \mu_3 > 0, \quad  ({\rm  rod \  \ molecule}),
\nonumber \\
&&\mu_1 < 0, \quad \mu_2=\mu_3 >0, \quad  ({\rm  disc \  \  molecule}),
\nonumber \\
&& \frac{1}{2}(\mu_1+\mu_3)> \mu_2, \quad ({\rm rod}{\rm-}{\rm like  \ \
molecule}), \nonumber \\
&&   \frac{1}{2}(\mu_1+\mu_3)< \mu_2, \quad ({\rm disc}{\rm-}{\rm
like  \ \
molecule}).
\end{eqnarray}

The expressions (\ref{QUANTEIG}) together with the classification
(\ref{PHASES}),
imply  that in the quantum case we  do not have at our disposal  either
spherically symmetric
or uniaxial molecules to begin with. The description used allows
only biaxial
molecules which are rod-like $ ( p > q)$  or disc-like $(p < q)$.

Using the parametrization (\ref{EIGPAR}) we obtain the
following
expressions for the functions of $p, q$ appearing in the  partition
function
(\ref{Z0EXPL})
\begin{eqnarray}
\label{CUANPAR}
&&-\frac{1}{3}(p-q)=\mu \ {\rm cos}(\psi-\frac{2 \pi}{3}), \quad
-\frac{1}{\sqrt{3}}(p+q+2)= \mu \ {\rm sin} (\psi-\frac{2 \pi}{3}),
\nonumber\\
&&-\frac{1}{3}( p+ 2q +3)=\mu \ {\rm cos}(\psi+\frac{2 \pi}{3}), \quad
\frac{1}{\sqrt{3}}(p+1)=\mu \ {\rm sin}(\psi+\frac{2 \pi}{3}),
\nonumber \\
&&+\frac{1}{3}( 2p + q +3)= \mu  \ {\rm cos}\psi, \quad
\frac{1}{\sqrt{3}}(q+1)=\mu \ {\rm sin}\psi.
\end{eqnarray}

\bigskip

\section{Parametrization of the equilibrium phases}

The  equilibrium state of the system is characterized by the vector
$B_a$ in
the frame where its only components are  $B_3$ and
$B_8$.
Again, an  $SU(3)$ invariant description of such phase
will be given by the corresponding vector invariants, in the adjoint
representation
\begin{equation}\label{INVAD}
{\bar I}_2= B_a B_a = B_3{}^2 + B_8{}^2, \quad {\bar I}_3=  d_{a b
c } B_a B_b
B_c =
\frac{1}{\sqrt{3}}( - B_8{}^3 + 3 B_8 B_3{}^2).
\end{equation}
 These invariants take continuous values as oppossed to those
describing the
individual molecules, which have  only discrete values.

 From Eqs (\ref{INVAD}), we can verify that $B_8$ satisfies the cubic
equation
\begin{equation}\label{CUBB}
B_8{}^3- {\bar I}_2 B_8 - \frac{2}{3} {\bar I}_3=0
\end{equation}
and consequently can be parametrized in any of the forms
(\ref{EIGPAR}). A
convenient choice is made by considering that the relevant variable
is the
matrix  $B_a \lambda_a$ in the fundamental representation,  which
eigenvalues
are
\begin{eqnarray}\label{EQEIG1}
&&{\vec \mu}_1 \cdot {\vec B}=\lambda_1= \frac{1}{2} B_3 +
\frac{1}{2\sqrt 3}
B_8:= \lambda \ {\rm cos} \left ( \phi + \frac{2 \pi}{3}\right),
 \nonumber \\
&& {\vec \mu}_2 \cdot {\vec B}=\lambda_2 = -\frac{1}{2} B_3 +
\frac{1}{2\sqrt
3} B_8:= \lambda \ {\rm cos} \left ( \phi - \frac{2 \pi}{3}\right),
\nonumber
\\
&& {\vec \mu}_3 \cdot {\vec B}=\lambda_3= - \frac{1}{\sqrt 3} B_8 :
= \lambda
\ {\rm cos} \ \phi .
\end{eqnarray}
In this way we identify
\begin{equation}\label{BEQPAR}
 B_8= -{\sqrt 3} \  \lambda \ {\rm cos} \ \phi, \qquad B_3= -{ \sqrt 3} \
\lambda \ {\rm sin } \phi.
\end{equation}
The equilibrium phase  is described by the variables $
\lambda, \phi$,
which can be translated in a geometrical language regarding the
average shape
of the equilibrium states
via the relations  (\ref{PHASES}) and (\ref{EQEIG1}),  in terms of the
eigenvalues
$\lambda_i$.

\bigskip

\section{Connection with the classical case}

The parametrization (\ref{CUANPAR}) allows us  to rewrite the
quantum partition
function in a form that closely resembles the classical case. In
fact, the
factor $N^{(p,q)}$  of $Z_0{}^{(p,q)}$, given in Eq. (\ref{NUM}),
can be rewritten as
\begin{equation}
\label{NUMPAR}
N^{(p,q)}=\frac{1}{2} A(X, \phi-\psi) -\frac{1}{2} A(X, \phi+\psi) =
\frac{1}{2}  D(X,  \phi, \psi),
\end{equation}
where $X= \frac{\sqrt{3}}{2} \mu y $ and $ D(X,  \phi, \psi)$ is the same
expression as in the classical case, since we have
\begin{equation}
\label{BASNUM}
A(X, \alpha)= {\rm e}^{X {\rm cos}\alpha}+{\rm e}^{X {\rm cos}(\alpha-
\frac{2 \pi}{3})} + {\rm e}^{X {\rm cos}(\alpha + \frac{2 \pi}{3})}.
\end{equation}
Here we have introduced the following dimensionless variables
\begin{equation}\label{FINVAR}
T= \frac{4}{\beta J}, \quad y=\frac{4\sqrt 3}{T} \  \lambda, \quad \rho=
-\frac{y}{2} {\rm sin} \ \phi,
\quad \sigma= -\frac{y}{2 \sqrt3}  {\rm cos } \ \phi.
\end{equation}
The temperatute $T_c$ used in Ref. \cite{MULDER} is related to our
definition by
\begin{equation}
\label{RELT}
T_c= \frac{1}{3 \mu^2} T,
\end{equation}
where $\mu$ is given by Eq.(\ref{PARCAS}). In the sequel we will refer to
$T_c$ as the classical temperature.

The denominator of the partition function can also be written as
\begin{equation}
\label{DENZ0}
N^{(0,0)}=- 4 \ {\rm sinh}( \frac{y}{2} {\rm sin}\phi ) \ {\rm sinh}(
\frac{y}{2} {\rm sin}(\phi- \frac{2 \pi}{3}) ) \ {\rm sinh}(
\frac{y}{2} {\rm
sin}(\phi + \frac{2\pi}{3})),
\end{equation}
which allows a more direct comparison with the classical case.

In
fact, in the
limit  $ T \rightarrow \infty \  ( y \rightarrow 0, X \rightarrow  
0) $ the
above expression reduces to
\begin{equation}
\label{LIMDENZO}
N^{(0,0)}=- \frac{1}{2} y^3 \ {\rm sin}\phi \ {\rm sin}(\phi-
\frac{2 \pi}{3})\
{\rm sin}(\phi + \frac{2 \pi}{3})= \frac{1}{8} y^3 \ {\rm sin}3 \phi.
\end{equation}
Since in the case $p=q=0$ we have \  $\mu = \frac{2}{\sqrt{3}}, \
y = X, \
I_3= I_2=0, \ {\rm sin} 3\psi =1,  $ we can write  $N^{(0,0)}=
\frac{1}{8} \
X^3 \  {\rm sin} 3\phi \  {\rm sin} 3\psi $. Then, in this limit we have
\begin{equation}\label{LIMZ0}
 Z_0^{(p,q)} |_{T  \rightarrow  \infty} = \frac{4 D(X, \phi, \psi  
)}{X^3 \
{\rm sin} 3
\phi \  {\rm sin} 3 \psi}.
\end{equation}

In order to compare the full quantum partition
function  with its classical counter part,
we  subsequently make a precise connection
between the two approaches, by rewriting the variable $x_c= \frac{3}{2}
\lambda_c \mu_c$ used in \cite{MULDER}, in our terms. Here $\mu_c$
denotes the
"radius"  in the parametrization of the
constituent phase eigenvalues, and $\mu_c=\mu_q$,  according to
Eqs.(\ref{QUANTEIG}). The subindices $c, q$ refer to  the classical
 \cite{MULDER} and quantum cases, respectively.  The value
$\lambda_c$ is  the
"radius" of the parametrization of the equilibrium phase, given by the
eigenvalues $\lambda_i$ in Ref \cite{MULDER}. The  variable ${\bar
S}_a$ of
Ref. \cite{MULDER} is what we call $B_a$ and the matrix whose
eigenvalues are
the $\lambda_i$'s is  $M_1= \frac{\beta J}{2 } B_a (2\lambda_a)$,
where we have
used that $({\lambda_a})_c = 2 (\lambda_a)_q$. We have labeled  the
"radius"
associated to the parametrization of  the eigenvalues of  $B_a
\lambda_a$ by
$ \lambda_q= \frac{y \  }{\sqrt{3}  \beta J }$ in such a way that $
{\lambda}_c={\beta J} \lambda_q $. Then we obtain $ x_c=\frac{3}{2}
\beta J
\lambda_q \mu_q $. On the other hand, we have that $X=
\frac{\sqrt{3}}{2} \mu_q
\  y = \frac{3}{2}  \beta J \lambda_q \mu_q  = x_c$. In this way,
(\ref{LIMZ0}) is
exactly the partition function given in  Ref \cite{MULDER}.

Thus, we can readily see that the numerator of  our quantum
partition function
(\ref{ZFRACC}) coincides  with the classical case, except
for the fact
that the angle $\psi$ is rectricted to discrete values according to the
relations  (\ref{CASREP1}), (\ref{CASREP2}), (\ref{PARCAS}),
(\ref{QUANTEIG})
and  (\ref{EIGPAR}). The denominator is quite different from the
classical case and shows no dependence upon the 
angle $\psi$, i.e. upon the input
parameters  $p, q$.

In general, we can relate the quantum ($Z_0{}^{(p,q)}$ ) and 
classical ($(Z_0)_c = Q $)
partition functions
by
\begin{equation}\label{RELZ}
 Z_0= {\cal Z } \ Q,
\end{equation}
with
\begin{eqnarray}\label{FACTOR}
{\cal Z} = D^{(p,q)}
\frac{\frac{ y}{2} {\rm sin} \phi}{{\rm sinh}(\frac{ y}{2} {\rm
sin} \phi)} \
\frac{\frac{ y}{2} {\rm sin}( \frac{2 \pi }{3} -\phi)}{{\rm
sinh}(\frac{ y}{2}
{\rm sin}(\frac{2 \pi}{3}-  \phi))} \  \frac{\frac{ y}{2} {\rm
sin}( \frac{2
\pi}{3} + \phi)}{{\rm sinh}(\frac{ y}{2} {\rm sin}( \frac{2 \pi
}{3}+ \phi))},
\end{eqnarray}
where $ D^{(p,q)}$ is de dimension of the $(p,q)$ representation of
$SU(3)$. Let us observe that the quantum calculation depends upon
the parameter $\mu$, as oppossed to the  
classical
case where this variable appears only through the combination $x_c$.

\bigskip

\section{Calculation of the equilibrium parameters}

In the  coordinates chosen so far, the consistency equations in the
mean field
approximation are given by
\begin{equation}\label{EQINSYS}
B_3= \frac{1}{\beta J} \frac{\partial}{\partial B_3} {\rm ln}
Z_0{}^{(p,q)},
\quad
B_8= \frac{1}{\beta J} \frac{\partial}{\partial B_8} {\rm ln}
Z_0{}^{(p,q)},
\end{equation}
where the partition function $Z_0{}^{(p,q)}$ is written in
Eq.(\ref{Z0EXPL}). The original expression
\begin{equation}
\label{ORFORM}
B_a=\frac{1}{Z_0} Tr \left( S_a e^{\beta J (B_3S_3 + B_8 S_8)}  
\right), \quad
a=3,8,
\end{equation}
for the consistency conditions, allows us to see that we will  
always have the
solution
$B_3=0=B_8, \  ( \lambda=  y =0)$ corresponding to the isotropic phase.
In this case, the partition function is $Z_0 =  
D^{(p,q)}$,
independent of  the variable $\phi$, which we choose as $\phi=0$ to  
describe
the isotropic phase. Nevertheless, this phase will not always be 
the stable one. 

Changing  to the variables $\lambda, \ \phi $ introduced in
Eq.(\ref{BEQPAR}),
the equilibrium equations reduce to
\begin{equation}\label{FEQEC}
\lambda= \frac{1}{3 \beta J} \frac{\partial}{\partial \lambda} \ {\rm ln}
Z_0{}^{(p,q)}, \qquad
\frac{\partial}{\partial \phi} \   {\rm ln} Z_0{}^{(p,q)}=0,
\end{equation}
which can be rewritten as
\begin{equation}\label{FINEQ}
\frac{1}{4} y  \ T=   \frac{\partial}{\partial y} \ {\rm ln}
Z_0{}^{(p,q)},
\quad
0= \frac{\partial}{\partial \phi} \ {\rm ln} Z_0{}^{(p,q)},
\end{equation}
in terms of the variables defined in Eq.(\ref{FINVAR}).

The free energy  $F$  of the system is
\begin{equation}\label{FREN}
\beta F = -{\rm ln} Z_0{}^{(p,q)}+\frac{T}{2}\rho^2 +\frac{3T}{2}
\sigma^2
=-{\rm ln} Z_0{}^{(p,q)} + \frac{1}{8} \ T  y^2,
\end{equation}
which has to be an absolute local minimum, according
to Eqs.
(\ref{FINEQ}), in order to determine the equilibrium state of the system.

The thermodynamic properties are given by
\begin{equation}\label{TERMO}
\frac{{\rm S}}{{\rm k}}= -\frac{1}{4} y ^2 \ T + ln Z_{0}{}^{(p,
q)}, \quad
 \frac{\rm U }{J}=-\frac{1}{32}  T^2 y^2, \quad
\frac{\rm C}{\rm k}= - \frac{1}{8} \frac{d}{d T} ( y^2 T^2),
\end{equation}
where $\rm S$ is the entropy per particle due to  the orientational
order,
${\rm U}$ is the  energy
per particle and  $\rm C$ is the specific heat. They  correspond to  
 the same
expressions
used  in  Ref.  \cite{MULDER}.

In order to study the symmetries of the partition function $
Z_0{}^{(p,q)}$  it
is convenient to rewrite it  as
\begin{equation}
\label{WEYL1}
Z_{0}{}^{(p, q)} =
\frac{1}{\left(\Lambda_{1}-\Lambda_{2}\right)
      \left(\Lambda_{1}-\Lambda_{3}\right)
      \left(\Lambda_{2}-\Lambda_{3}\right)}
\left|\begin{array}{ccc}
   \Lambda_{1}^{-(q+1)} & \Lambda_{1}^{(p+1)} & 1  \\
   \Lambda_{2}^{-(q+1)} & \Lambda_{2}^{(p+1)} & 1  \\
   \Lambda_{3}^{-(q+1)} & \Lambda_{3}^{(p+1)} & 1
\end{array} \right|,
\end{equation}
together with
\begin{equation}\label{LAMDAS}
\Lambda_1= {\rm e}^{\frac{1}{\sqrt 3}  \  y {\rm cos} \left( \phi +
\frac{2
\pi}{3} \right)} , \quad
\Lambda_2= {\rm e}^{\frac{1}{\sqrt 3}  \ y  {\rm cos} \left( \phi -
\frac{2
\pi}{3} \right)} , \quad
\Lambda_3= {\rm e}^{\frac{1}{\sqrt 3}  \ y  {\rm cos}  \phi },
\end{equation}
according to Eqs. (\ref{EIGEN}), (\ref{EQEIG1}) and (\ref{FINVAR}).
The partition function is invariant under each of the following
separate set of
 transformations, which do not change the system under consideration
 \begin{eqnarray}
&& y\rightarrow y, \quad \phi\rightarrow- \phi, \quad \quad
\left( \Lambda_1\leftrightarrow \Lambda_2{}, \quad \Lambda_3
\rightarrow
\Lambda_3\right), \label{SU1}\\
&& y\rightarrow y, \quad \phi\rightarrow \frac{2 \pi}{3}-\phi, \quad
\left( \Lambda_3\leftrightarrow \Lambda_2{}, \quad \Lambda_1
\rightarrow
\Lambda_1\right), \label{SU2}\\
&&  y\rightarrow y, \quad \phi\rightarrow \frac{4\pi}{3}-\phi, \quad
\left( \Lambda_3\leftrightarrow \Lambda_1{}, \quad \Lambda_2
\rightarrow
\Lambda_2\right) \\
&&  y\rightarrow y, \quad \phi\rightarrow \phi-\frac{2\pi}{3}, \quad
\left( \Lambda_1\rightarrow \Lambda_3 \rightarrow \Lambda_2
\rightarrow
\Lambda_1\right) \\
&&  y\rightarrow y, \quad \phi\rightarrow \phi+\frac{2\pi}{3}, \quad
\left( \Lambda_1\rightarrow \Lambda_2 \rightarrow \Lambda_3
\rightarrow
\Lambda_1\right).\label{SIM1}
\end{eqnarray}

The above set of transformations allow us to restrict the initial
interval $- \pi < \phi < \pi$ to $0< \phi < \frac{\pi}{3}$.
Furthermore, we
have the following relations among the different values
of $\Lambda_i$
\begin{eqnarray}\label{RELLAM}
&&0< \Lambda_1 < \Lambda_2 < 1 < \Lambda_3, \quad
0< \phi < \frac{\pi}{6},\nonumber  \\
&&0< \Lambda_1 < 1 < \Lambda_2 < \Lambda_3, \quad
\frac{\pi}{6} < \phi < \frac{\pi}{3}.
\end{eqnarray}
The remaining symmetries exchange rod-like into disc-like phases
\begin{eqnarray}
&&p \leftrightarrow q, \quad  y\rightarrow y, \quad \phi\rightarrow
\pi- \phi,
\quad
\left( \Lambda_1\leftrightarrow \Lambda_2{}^{-1}, \quad \Lambda_3
\rightarrow
\Lambda_3{}^{-1}\right), \label{S1} \\
&&p \leftrightarrow q, \quad   y\rightarrow y, \quad \phi\rightarrow
\frac{\pi}{3}- \phi, \quad
\left( \Lambda_1\leftrightarrow \Lambda_3{}^{-1}, \quad \Lambda_2
\rightarrow
\Lambda_2{}^{-1}\right), \label{S2} \\
&&p \leftrightarrow q, \quad  y\rightarrow y, \quad \phi\rightarrow
\frac{\pi}{3}+ \phi, \quad
\left( \Lambda_2\leftrightarrow \Lambda_3{}^{-1}, \quad \Lambda_1
\rightarrow
\Lambda_1{}^{-1}\right). \label{S3}
\end{eqnarray}
The second of the above symmetries was already discussed in the
classical case of
 Ref. \cite{MULDER}.

For future purposes it is convenient to rewrite the ratio of
the different values of $\Lambda_i$ in the following way
\begin{equation}\label{RATIOS}
\frac{\Lambda_1}{\Lambda_3}= {\rm e}^{-y \ {\rm sin}( \phi +
\frac{\pi}{3})},
\quad
\frac{\Lambda_2}{\Lambda_3}= {\rm e}^{-y \ {\rm sin}( \frac{\pi}{3}
- \phi)}, \quad
\frac{\Lambda_1}{\Lambda_2}= {\rm e}^{-y \ {\rm sin}\phi}, \quad
\end{equation}
%%%%%%%%%%%%%%%%%%%%%%%%%%%%%%%%%%%%%%%%%%%
%%%%%%%%%aqui termina fin1.tex%%%%%%%%%%%%%%%%%%%%%
%%%%%%%%%%%%%%%%%%%%%%%%%%%%%%%%%%%%%%%%%

%%%%%%%%%%%%%%%%%%%%%%%%%%%%%%%%%%%%%%%%%%%
%%%%%%%%%aqui termina fin1.tex%%%%%%%%%%%%%%%%%%%%%
%%%%%%%%%%%%%%%%%%%%%%%%%%%%%%%%%%%%%%%%%

\section{The phases of the system}

 The free energy  $\beta F$ of the system, given by Eq.  (\ref{FREN}),
is a function of
five variables; $\beta F=\beta F(T, p, q, y, \phi)$. The absolute minimum
conditions (\ref{FINEQ}), which correspond to two equations in
our case, allow us to  find $y=y(T,p,q)$ and $\phi=\phi(T,p,q)$ for
equilibrium, thus selecting the particular phase which is
energetically favourable.  This calculation is made numerically,  
and these
results
determine all the remaining thermodynamic properties of the system.
The
entropy, energy and specific heat can be subsequently calculated by 
using the expressions (\ref{TERMO}). The general behavior of the
specific heat, as a function of temperature is presented in  Fig. 1.
Here we have labelled  by $T_b$, $T_u$, the transition temperatures
between the biaxial-uniaxial phase and the uniaxial-isotropic phase,
respectively. Since we are only considering the description of the
orientational  modes of the system, the thermodynamic quantities  
associated
with the isotropic phase will all be zero.

\subsection{The T $\rightarrow 0  \ $(y $\rightarrow \infty$) case.}

Let us consider the interval $ 0 < \phi < \frac{\pi}{3}$
in such a way that all the ratios $(\ref{RATIOS})$ tend to zero
exponentially. The corresponding limit in the partition function
is taken from $(\ref{ZGEN})$, by observing that each of the polynomials
$F_1$ can be approximated as
\begin{equation}
\label{APPR}
F_1(q; \Lambda_2, \Lambda_3)\approx
{\Lambda_3}^q \left(1 + \frac{\Lambda_2}{\Lambda_3} \right).
\end{equation}
In this way, the product  of two F's in each term of the partition
function factors out, leaving an additional term
$F_1(q; \Lambda_1, \Lambda_2)$, which is also approximated in the
way of Eq.(\ref{APPR}). The final expression for the logarithm of
the partition
function
is
\begin{equation}
\label{ZAPPROX}
{\rm ln} Z_0^{(p,q)}= (p+q){\rm ln} \Lambda_3 + q {\rm ln} \Lambda_2 +
\frac{\Lambda_1}{\Lambda_3}+ \frac{\Lambda_2}{\Lambda_3}+
\frac{\Lambda_1}{\Lambda_2}.
\end{equation}
In order to determine which of the last three terms in the above
expression dominates, we still have to consider the  further intervals
$0< \phi \ < \frac{\pi}{6}$, where ${\rm sin}\phi < {\rm sin}(\frac{\pi}
{3}- \phi) < {\rm sin}( \frac{\pi}{3} + \phi)$, together with
$\frac{\pi}{6} < \phi \ < \frac{\pi}{3}$, where $ {\rm sin}(\frac{\pi}
{3}- \phi)< {\rm sin}\phi  < {\rm sin}( \frac{\pi}{3} + \phi)$. In the
first case, the free energy reduces to
\begin{equation}
\label{AP1}
\beta F= -(p+q)\frac{y}{\sqrt{3}}{\rm cos}\phi - q
\frac{y}{\sqrt{3}}{\rm cos}(\phi- \frac{ 2\pi}{3})
- {\rm e}^{- y {\rm sin}\phi} + \frac{1}{8} T y^2,
\end{equation}
because the term $\frac{\Lambda_1}{\Lambda_2}$ dominates over the  
remaining
fractions in (\ref{ZAPPROX}).
Introducing the notation
\begin{eqnarray}
\label{NOTAP1}
&A(\phi)=\frac{p+q}{\sqrt{3}}{\rm cos}\phi
+ \frac{q}{\sqrt{3}}{\rm cos}(\phi- \frac{ 2\pi}{3}),\nonumber \\
&B(\phi)=\frac{p+q}{\sqrt{3}}{\rm sin}\phi +
\frac{q}{\sqrt{3}}{\rm sin}(\phi- \frac{ 2\pi}{3}),
\end{eqnarray}
the equations that minimize the free energy are
\begin{eqnarray}
\label{EQAP1}
&-A(\phi)+ \frac{1}{4}T y  + {\rm sin}\phi \ {\rm e}^{-y {\rm
sin}\phi}=0,
\nonumber \\
&B(\phi) + {\rm cos}\phi \ {\rm e}^{-y {\rm sin}\phi}=0.
\end{eqnarray}
The above equations are solved by
\begin{eqnarray}
\label{SOLAP1}
&\phi=\phi_0 - \frac{{\rm cos}\phi_0}{A(\phi_0)}{\rm e}^{-\frac{4
A(\phi_0)}{T}{\rm sin}\phi_0}, \quad
{\rm tan}\phi_0=\frac{{\sqrt 3}q}{2p +q},\nonumber \\
&y=\frac{4}{T}\left( A(\phi_0)- {\rm sin}\phi_0
 \ {\rm e}^{-\frac{4 A(\phi_0)}{T}{\rm sin}\phi_0} \right).
\end{eqnarray}
The solution in the interval $\frac{\pi}{6} <\phi < \frac{\pi}{3}$
is
\begin{eqnarray}
\label{SOLAP2}
&\phi=\phi_0 - \frac{{\rm cos}(\frac{\pi}{3}-\phi_0)}{A(\phi_0)}{\rm
e}^{-\frac{4 A(\phi_0)}{T}{\rm sin}(\frac{\pi}{3}-\phi_0)}, \quad
{\rm tan}\phi_0=\frac{{\sqrt 3}q}{2p +q},\nonumber \\
&y=\frac{4}{T}\left( A(\phi_0)- {\rm sin}(\frac{\pi}{3}-\phi_0)
 \ {\rm e}^{-\frac{4 A(\phi_0)}{T}{\rm sin}(\frac{\pi}{3}-\phi_0)}
\right).
\end{eqnarray}

This limit clearly corresponds to the biaxial equilibrium phase since
$\lambda \neq 0, \phi \neq 0$. For the case $0 < \phi < \frac{\pi}{6}$,
we obtain the following expression for the specific heat
\begin{equation}
\label{CT0}
\frac{{\rm C}}{{\rm k}}=\left(\frac{2 q}{T}\right)^2
{\rm e}^{- \frac{2 q}{T}},
\end{equation}
where there is a  mass gap  given by $2 q$.
The prediction (\ref{CT0}) has to be compared with the classical case,
where the specific heat goes like $\frac{{\rm C}}{{\rm k}}=3 +
\frac{3T}{( \mu  \ {\rm sin} 3 \psi)^2}$ in that limit.  Moreover,
Eq.(\ref{CT0})
signals a peculiar behavior for the cases $q=0, \  p \neq 0$, which
will be separately discussed in section 7.4.

The solution (\ref{SOLAP2}) in the interval $\frac{\pi}{6} \leq  
\phi  \leq
\frac{\pi}{3}$ leads to
\begin{equation}
\label{CT01}
\frac{{\rm C}}{{\rm k}}=\left(\frac{2 p}{T}\right)^2
{\rm e}^{- \frac{2 p}{T}},
\end{equation}
for the specific heat in the $T \rightarrow 0$ limit. This is  
consistent with
the symmetry
$p \leftrightarrow q$ which connects the corresponding  two  
intervals of $\phi$.

\subsection{Biaxial-Uniaxial phase transition}

The uniaxial phase is characterized by $\lambda \neq 0, \phi =0$.
In order to
determine the transition temperatute $T_b$, we expand the free energy
$F(T, \lambda, \phi)$ in powers of $\phi$ and find the
corresponding minimum.
The expansion is
\begin{equation}
\label{BUEXP}
\beta F(T, \lambda, \phi)=  f(z)- 6 z +
6 \ T z^2 + (z^2 g(z) + z \ h(z))\phi^2 + {\rm O}(\phi^4),
\end{equation}
where the functions $f, g, h $ are  explicitly known and
$z=\frac{\lambda}{T}$. The partition function is invariant under
the change
$\phi \rightarrow -\phi$, which is reflected in the above expansion. To
determine the transition temperature we approach from the biaxial
phase and
look for the corresponding minimum of (\ref{BUEXP}) having $
\lambda \neq 0, \
\phi \neq 0,   \ {\rm  but \ small}$. The conditions are
\begin{equation}\label{MIN1BU}
0= \frac{\partial F}{\partial \phi}=2 \phi z \left[ \lambda g(z) + T h(z)
\right] + {\rm O}(\phi^3),
\end{equation}
\begin{equation}\label{MIN2BU}
0= \frac{\partial F}{\partial \lambda}= \frac{\partial
f(z)}{\partial z}-6
+12 \lambda +{\rm O}(\phi^2).
\end{equation}
 From Eq.(\ref{MIN1BU}) we obtain ${\bar z} g(\bar z) + h(\bar
z)=0$, which
numerical solution determines $ \bar z= \frac{\lambda_b}{T_b}$.
Substituting
this result in Eq.(\ref{MIN2BU}) the transition temperature results in
\begin{equation}\label{TTBU}
T_b=\frac{1}{12 \bar z}\left(6 - \frac{\partial f(\bar z)}{\partial
\bar z}
\right).
\end{equation}
Some  results, for different values of $p$  
and $q$, are presented in Table I. In this table, the
parameters   $\mu$  and $\psi$  
corresponding
to the quantum numbers $p$ and $q$ are obtained from Eqs.  
(\ref{CASREP1}),
(\ref{CASREP2}) and
(\ref{PARCAS}) The fifth column is obtained from Eq. (\ref{TTBU}).  
The next
columm shows the conversion to the classical temperature, according to  
Eq.
(\ref{RELT}). The predictions of the classical model
are  presented in the last  
columm.

\subsection{Uniaxial-Isotropic phase transition}

As we keep increasing the temperature we go from the uniaxial phase
to the isotropic phase characterized by $\phi=0, \lambda=0$. To determine
the transition temperature $T_u$ we proceed in complete analogy to the
previous section. Now we expand the free energy of the uniaxial phase
in the vicinity of $z=0$,
\begin{eqnarray}\label{FEL0}
&&\beta F(z, \phi=0, T)=- {\rm ln}Z_0{}^{(p,q)}+\frac{3}{2}T
\sigma^2= 6
\ T z^2 -3I_2 z^2 - \frac{12}{5}I_3 z^3\nonumber \\
&&-{\rm ln}\frac{(p+1)(q+1)(p+q+2)}{2} + \frac{9}{10}I_2(I_2+2) z^4 +
\frac{108}{35}I_2 I_3 z^5 + {\rm O}(z^6), \nonumber \\
\end{eqnarray}
where we recall that  $z=\frac{\lambda}{{T}}$.

The extremum condition $\frac{\partial F}{\partial \sigma}=0$
requires $3 T \sigma= \frac{\partial {\rm ln}
Z_0{}^{(p,q)}}{\partial \sigma}$.
The general form of $\frac{\partial {\rm ln} Z_0{}^{(p,q)}}{\partial
\sigma}$ as
a function of $\sigma$ is presented in  Fig. 2. 
From here it is clear that the transition temperature
is given by the slope of the straight line which is tangent to this
curve,
i.e. there is a critical temperature beyond which it is
not possible to satisfy the
consistency conditions. This implies the further condition $3 T=
\frac{\partial^2 {\rm ln} Z_0{}^{(p,q)}}{\partial \sigma ^2}$. Both
requirements imply that one is looking for an inflexion point
of the free energy (\ref{FEL0}): $\frac{\partial F}{\partial  
\sigma}=0, \;\;
\frac{\partial^2 F}{\partial \sigma^2}=0$.

Keeping up to fourth order terms in (\ref{FEL0}), the above system
is solved by
\begin{equation}
\label{TAST}
z^*=\frac{I_3}{I_2(I_2+2)}, \quad T^*=\frac{1}{2}I_2 +\frac{3}{10}
\frac{I_3^2}{I_2(I_2+2)}.
\end{equation}
This solution is a good aproximation whenever $z^*$ is small.

Nevertheless, for a range of temperatures below $T^*$, we can verify that
the uniaxial phase does not provide an absolute minimum for the free
energy of the system. In fact, the calculated free energy is higher
than the
one corresponding to the isotropic phase, which makes the uniaxial phase
metastable. In this way, the true
transition temperature $T_u$ between the uniaxial and the isotropic
phase is
obtained by the minimum condition together with the requiremente that
$\beta F= \beta F|_{\rm isotropic}=-{\rm
ln}\frac{1}{2}(p+1)(q+1)(p+q+2)$.
In the same approximation as the one used in Eq.(\ref{TAST}), these new
conditions provide
\begin{equation}
\label{TU}
z_u=\frac{4 I_3}{ 3 I_2(I_2+2)}, \quad T_u=\frac{1}{2}I_2 +\frac{4}{15}
\frac{I_3^2}{I_2(I_2+2)},
\end{equation}
which confirm that $T_u < T^*$ together with the fact that the metastable
region is small indeed. A further discussion of the metastability  
region can be
found in section 4.5 of Ref. \cite{CHAILUB}

We  can calculate the limit
of the specific heat when we approach $T_u$ fom the left, in  the  
case when
$I_3\rightarrow 0 \ (p \rightarrow q)$. To this end we have 
to find $z=z(T)$, which we  
obtain by requiring
$\frac{\partial F}{\partial z}=0$ in  Eq.(\ref{FEL0}). We get
\begin{equation}
\label{ZT}
z=\frac{1}{I_2(I_2+2)}\left( I_3 +
\sqrt{I_3^2+ \frac{10}{3}I_2(I_2+2)( \frac{I_2}{2}-T)} \right).
\end{equation}
{}From the above equation we can verify that if we substitute
the value for $T_u$ given by Eq.(\ref{TU}), we recover the value $z_u$
written in the same equation. Next we use the expression (\ref{TERMO})
for the specific heat to obtain
\begin{equation}
\label{CTU}
\frac{{\rm C}}{{\rm k}}{|}_{T \rightarrow T_u}= 20 \frac{I_2}{I_2+2}
+ O(I_3^2).
\end{equation}
This expression has the correct classical limit given by
$\frac{{\rm C}}{{\rm k}}{|}_{T \rightarrow T_u}= 20$.

The above results (\ref{TAST}), (\ref{TU}) can be further refined
by taking into account the term $z^5$ en  the expression  
(\ref{FEL0}). Some
numerical  results
are presented in   Table II, for different values of $p$ and $q$.
The fourth  columm gives the quantum prediction for the transition  
temperature
$T_u$, which is subsequently
converted to the classical temperature  in  the fifth columm. The last columm  
contains
the classical predictions.

\subsection{ The $q=0, p\neq 0$ case}

As we will see in the sequel, this situation has no counterpart
in the classical description  because here we have  only an isotropic-uniaxial
transition
and the biaxial phase is not present.

Let us consider the limit $T\rightarrow 0$ within the range
$0 \leq \phi \leq \frac{\pi}{6}$ in the expression (\ref{Zpq0})
for the exact partition function. As before, $\Lambda_1,
\Lambda_2 \rightarrow 0$ while $\Lambda_3 \rightarrow \infty$ in  
such a way

that the expansion of the partition function up to second order
terms leads to
\begin{equation}
{\rm ln}Z_0^{(p, q=0)}=p  \ {\rm ln}\Lambda_3 +  
\frac{\Lambda_1}{\Lambda_3}
+ \frac{\Lambda_2}{\Lambda_3} -  
\frac{1}{2}\left(\frac{\Lambda_1}{\Lambda_3}
\right)^2 -\frac{1}{2}
\left( \frac{\Lambda_2}{\Lambda_3} \right)^2 + \dots.
\label{Zpq0AP}
\end{equation}

The above expression differs from the $q=0$ restriction of Eq.
(\ref{ZAPPROX}), in the absence of the previously dominating
term $\frac{\Lambda_1}{\Lambda_2}$. This is because one has  to  
take the $q=0$
value in the full expression and not in a truncation of it. For
example, $F_1(q=0 ; u,v)=1$, but in an approximation where
$u \geq v$, we can write $F_1 =u^q ( 1 + \frac{v}{u} + \dots)$, which
will lead to $F_1(q=0 ; u,v)=( 1 + \frac{v}{u} + \dots)$. It is enough to
consider the expression (\ref{Zpq0AP}) to first order, which
leads to
\begin{equation}
\beta F=-\frac{p}{\sqrt{3}} y {\rm cos}\phi -
{\rm e}^{- y {\rm sin}(\frac{\pi}{3}-\phi)} -
{\rm e}^{- y {\rm sin}(\frac{\pi}{3}+\phi)} +\frac{1}{8} T y^2.
\label{Zpq0AP1}
\end{equation}
Since we are interested in solutions near $\phi=0$, we have to keep
both exponentials, which are of the same order now. The conditions for a
minimum of the free  energy are
\begin{equation}
\label{pq0EQ1}
\frac{1}{4} T y= \frac{p}{ \sqrt{3}} {\rm cos}\phi -
{\rm sin}(\frac{\pi}{3} - \phi)
{\rm e}^{- y {\rm sin}(\frac{\pi}{3}-\phi)}-
{\rm sin}(\frac{\pi}{3}+\phi)
{\rm e}^{- y {\rm sin}(\frac{\pi}{3}+\phi)}
\end{equation}
\begin{equation}
\label{pq0EQ2}
{\rm sin}\phi=\frac{\sqrt{3}}{p}\left({\rm cos}(\frac{\pi}{3}+\phi)
{\rm e}^{- y {\rm sin}(\frac{\pi}{3}+\phi)}
-{\rm cos}(\frac{\pi}{3}-\phi)
{\rm e}^{- y {\rm sin}(\frac{\pi}{3}-\phi)}\right):= H(\phi).
\end{equation}
In writing the last equation we are assuming that $y\neq0$. We
have that $H(\phi=0)=0$ and it is a direct matter to show that
$\frac{d H}{d \phi} <0$, within  the interval under consideration.  
This means
that the only solution to Eq.(\ref{pq0EQ2}) is $\phi=0$,
i.e. we remain in  the uniaxial phase for low temperatures, with no  
transition
to the biaxial one. The remaining equilibrium
condition is
\begin{equation}
\label{pq=0y}
 y=\frac{4}{T \sqrt{3}}\left( p- 3 {\rm e}^{-\frac{2p}{T}} \right).
\end{equation}
The above expression leads to a specific heat given by
\begin{equation}
\frac{\rm C}{\rm k}=\frac{8 p^2}{T^2} {\rm e}^{-\frac{2p}{T}},
\end{equation}
in the $T\rightarrow 0, y \rightarrow \infty$ limit.

The absence of the biaxial phase in this case can also be verified
from the general discussion of section 7.2, by looking at the  
expansion of the
free energy in powers of $\phi$. It is possible to
verify numerically that the coefficient of the power $\phi^2$ in Eq.
(\ref{BUEXP}) cannot be made zero, unless $z=0$. That is to say,  
there can be
no transition to the biaxial phase when $q=0, \ p \neq0$.

Another piece of information related to the above statement is
the calculation of the angle $\psi$. The general result is
\begin{equation}
\label{PSI}
{\rm cos} 3 \psi= \frac{p(p+3)(2p +3)}{2(p^2+ 3p +3)^{\frac{3}{2}}}.
\end{equation}
Now, in the limit $ p \rightarrow \infty$, we  recover the  
classical regime
together with the fact that ${\rm cos}3\psi=1$, i. e.  $\psi=0$.
Recalling the phase diagram of Ref.\cite{MULDER}, we verify that
in this regime we have only the uniaxial and isotropic phases.

\subsection{The $p=q$ case}

Here $I_2= p(p+2), \ I_3=0$. The specific heat, together with the entropy
are shown in Figs. 5 and 8, as a function of temperature. All the points correspond
to the biaxial phase and, exactly as in the 
classical situation, the phase transition is directly to the isotropic one. 
In order to better understand the limiting values of the transition temperature,
together with the  resulting specific heats, we we will take the corresponding limit starting  from the uniaxial phase, with $I_3=0$. In this case Eq.
(\ref{ZT}), valid for any point in this phase,  reduces to
\begin{equation}
\label{ZT30}
z= \sqrt{\frac{10}{3 \ I_2(I_2+2)}(\frac{I_2}{2}-T)}.
\end{equation} 
From Eqs.(\ref{TU}) we see that, in this limit, the uniaxial phase merges into the 
isotropic one, with $z=0$, at the transition temperature 
$T_b=T_u= {\tilde T}= \frac{I_2}{2}$.In the conventions of Ref. \cite{MULDER}, this
temperature
corresponds to 
\begin{equation}
{\tilde T}_c=\frac{1}{8}\frac{I_2}{I_2+1},
\end{equation}
which reproduces the value ${\tilde T}_c=\frac{1}{8}$ for high quantum numbers. The specifc heat at the transition is 
\begin{equation}
\frac{{\rm C}}{{\rm k}}|_{\tilde T}= 5 \  \frac{I_2}{I_2+2}=5  \ \frac{(p+1)^2-1}{(p+1)^2 +1}.
\end{equation}
The above result reproduces the classical prediction. 

Let us observe that Eq.(\ref{ZT30}) makes sense only when $T\leq \frac{I_2}{2}$. In other words, we really approach the transition point
from the sector of the uniaxial phase which has the biaxial phase as a boundary.

\subsection{ The order parameters of the system}

In the classical case, the order parameters corresponding to the  
uniaxial and
biaxial phases, respectively,  are defined by   \cite{MULDER}
\begin{equation}
\label{OPCL}
U=Tx \frac{\cos \phi}{\cos \psi}=\frac{2}{\beta J}\frac{\lambda_{c}
\cos \phi}{\mu \cos \psi},  \qquad
B=Tx \frac{\sin \phi}{\sin \psi}=\frac{2}{\beta J}\frac{\lambda_{c}
\sin
\phi}{\mu \sin \psi},
\end{equation}
where $\lambda_c$ refers to the notation in  this reference.
The  above definitions are such that $U \neq 0$, when $ T \leq T_u$  
and $B \neq
0$ when
$T \leq T_b$.

Considering that
\begin{equation}
\label{RELCQ}
\rho=-\frac{2 \sqrt{3}}{T} \lambda \sin \phi \;\;\;\;\;\;\;\;
\sigma=-\frac{2}{T} \lambda \cos \phi ,
\end{equation}
with  $ T=\frac{4}{\beta J} $ and   $ \lambda_{c}=\beta J  
\lambda $, the
corresponding order parameters in the quantum case can be taken as
\begin{equation}
\label{OPQU}
U= -\frac{T \sigma}{\mu \cos \psi}, \qquad  B=-\frac{T
\rho}{\sqrt{3} \mu \sin \psi},
\end{equation}
in order to reproduce the classical quantities in the apropriate  
limit. Let us
recall that
$\mu$ and  $\psi$  are related to $ p$  and $q $ through the equations
(\ref{CASREP1}),
(\ref{CASREP2})  and  (\ref{PARCAS}). Two complementary examples are 
presented in Figs. 10 and 11. The former corresponds practically 
to the classical situation, while the latter refers to a lower lying
quantum state.

\section{Final Comments and Conclusions}

We have presented a quantum mechanical description of  the orientational
degrees of freedom of a biaxial nematic liquid in the mean field  
approximation,
thus extending  the classical  approach proposed in  Ref. \cite{MULDER}.

The fact that the dynamical variables  carry a representation of  $SU(3)$
allows  for an exact calculation  of the quantum partition  
function, in terms
of the  Weyl formula for the characters of the  group. In this way, the
consistence equations of the mean field approximation
reduce to two coupled algebraic equations, in a manner similar to  
the case in
Ref. \cite{MULDER}.The work presented here constitute a nice example
of the applications of group theory techniques to quantum statistical
mechanics.
As expected \cite{LIEB, SIMON}, the classical behavior is recovered in the high  
temperature, high
quantum numbers regime.  In general, the discrepancies between the two
approaches begin to  
appear  at
temperatures  which are lower that the biaxial-uniaxial transition  
temperature.

Under these circumstances and considering the actual state of the art
in the production of liquid crystals, the liquid will
probably crystalize and these effects will be irrelevant from the  
experimental
point of view, in the case of  the nematic phase  under present-day
laboratory conditions.
This emphasizes  the fact that the classical approximation is very adequate  
for the
description of the phase transitions in this model. Nevertheless, further
technical developments may lead to nematic liquids with substantially lower
transition temperatures, such that quantum effects might turn to be important.
Also, it is necessary to continue exploring the relevance and description 
of liquid crystals in high density-high temperature nuclear matter.

The general phase structure of the system includes: a biaxial phase
in the range $0 < T < T_b$, a uniaxial phase in the range $ T_b < T < T_u
, \ T_b < T_u$,  and an isotropic phase for $T> T_u$. This last phase acts
here as a reference, because the model does not include a description of 
translational degrees of freedom of the liquid. This structure is exhibited
in  Figs. 3, 4 and 5 (for the specfic heat) together with Figs. 7 and 9 (for
the entropy). The specific heat always presents a discontinuity 
at $T_b$, while the
entropy remains continuous there. This corresponds to a second order phase
transition. We expect the same behavior at the temperature $T_u$, since
both phase transitions are associated to symmetry changes in the liquid.
The  temperatures of the different phase transitions in the
quantum calculation are about   
$15-30 \ \%$ lower than the classical situation, 
in the most unfavourable cases (low quantum
numbers). 

In the low temperature limit $T\rightarrow 0$,  the quantum  
calculation of the
specific heat
leads to the result $\frac{{\rm C}}{{\rm k}} \sim \frac{1}{T^2}  
{\rm e}^{-
\frac{\alpha}{T}}$ , according to Eqs. (\ref{CT0}) and  
(\ref{CT01}), signaling
a mass gap characteristic of the quantum spectra. The classical  
limit in this
regime is $\frac{{\rm C}}{{\rm k}} =3$. Also, the quantum calculation
is consistent with the Nernst postulate for the entropy in the $T\rightarrow
0$ limit.

Deviations from the above general picture are present in two cases: (i) the
region $p=q$ produces a direct transition from the biaxial to the isotropic
phase, in complete analogy with the classical situation. This is shown in Figs.
5 and 8.  (ii) the region
$q=0, \ p \neq 0 \ (p=0, \ q \neq 0)$ does not support the biaxial 
phase and only allows for
a uniaxial-isotropic phase transition. This whole 
quantum region is the analogous to the $\psi=0, (\psi = \frac{\pi}{3})$
limit in the classical case. These values for $\psi$ are not allowed in the
quantum regime, leading to $ q=-1 \ (p=-1)$.

Our ultimate goal, which is not discussed in this work,  would be to
incorporate  spin (magnetic moment)
degrees of
freedom into the
model for the nematic liquid, in order to study how the
thermodynamic phases
and  equilibrium parameters are modified .  Following an analogous
line of
development, this would require
the introduction of a supergroup as a way of  characterizing the
underlying orientational
degrees of freedom.
In this case  it would be also  interesting to  study  the
classical limit of
the quantum version. This limit will shed some light into the
 role of the supersymmetric extension \cite{SUSYIZ}   of the   
Harish-Chandra,
Itzykson - Zuber
 (HCIZ)  integral,  in the classical description of the model. A direct
classical calculation of the supersymmetric model, along the lines of
Ref.\cite{MULDER}, presents some difficulties, even though the  
corresponding
HCIZ integral is known  \cite{SUSYIZ} . In fact, there is no Macfarlane's
theorem in this case and the interpretation of the Grassmann numbers that
naturally arise is not at all clear. In this situation,  we expect  
the quantum
calculation to be free of  these ambiguities, besides of providing  
the complete
and correct  answer to the problem.

\vskip.5cm

\noindent
{\bf Acknowledgements}

\vskip.3cm

OCG acknowledges  support  of  Direcci\'on de Investigaci\'on y Postgrado
at  Universidad Cat\'olica de Chile (DIPUC) through the scholarship  
awarded to
him. LFU was supported in part by the projects CONACyT-CONICyT  
E120-1810 and
DGAPA-UNAM-IN100397. The work of JA  is partially supported by Fondecyt 1980806
and the CONACyT-CONICyT project 1997-02-038.
The authors also thank  
Professor  B.M.
Mulder for calling
their attention to the application of the HCIZ integral to the  
description of
nematic liquids. Finally
we thank  R. Benguria and N. Valdebenito for useful  
information and
suggestions.

\bigskip
\eject

{\LARGE  The Appendix}

\vskip.5cm

Any  set of hermitian operators $S_a,  \ a=1,2, \dots, 8$ satisfiying the
commutators
\begin{equation}\label{SU3}
[ S_a, \ S_b]= i f_{a b c} S_c,
\end{equation}
with $ f_{a b c}$ being the $SU(3)$ structure constants  carry a
representation
of the group

The fundamental representation of $SU(3)$ is provided by the Gell-Mann
$\lambda_a$ matrices

\bigskip

$\lambda_{1}= \frac{1}{2}\left(\begin{array}{ccc}
            0 & 1 & 0 \\
            1 & 0 & 0 \\
            0 & 0 & 0
         \end{array}
  \right), $
 $\lambda_{2}= \frac{1}{2}\left(\begin{array}{ccc}
            0 & -i & 0 \\
            i & 0 & 0 \\
            0 & 0 & 0
         \end{array}
  \right), $
 $\lambda_{3}= \frac{1}{2}\left(\begin{array}{ccc}
            1 & 0 & 0 \\
            0 & -1 & 0 \\
            0 & 0 & 0
         \end{array}
  \right), $

  \bigskip
 $\lambda_{4}= \frac{1}{2}\left(\begin{array}{ccc}
            0 & 0 & 1 \\
            0 & 0 & 0 \\
            1 & 0 & 0
         \end{array}
  \right), $
 $\lambda_{5}= \frac{1}{2}\left(\begin{array}{ccc}
            0 & 0 & -i \\
            0 & 0 & 0 \\
            i & 0 & 0
         \end{array}
  \right), $
  $\lambda_{6}= \frac{1}{2}\left(\begin{array}{ccc}
            0 & 0 & 0 \\
            0 & 0 & 1 \\
            0 & 1 & 0
         \end{array}
  \right), $

  \bigskip
 $\lambda_{7}= \frac{1}{2}\left(\begin{array}{ccc}
            0 & 0 & 0 \\
            0 & 0 & -i \\
            0 & i & 0
         \end{array}
  \right), $
 $\lambda_{8}= \frac{1}{2\sqrt{3}}\left(\begin{array}{ccc}
            1 & 0 & 0 \\
            0 & 1 & 0 \\
            0 & 0 & -2
         \end{array}
  \right), $

\bigskip
\noindent
 which are traceless and  satisfy  $[\lambda_{a},\lambda_{b}]=if_{a b c}
\lambda_{c}$. The explicit expression for the  $SU(3)$ structure
constants is
given by
\begin{equation}\label{ESTRC}
f_{a b c }= - 2 i tr ([ \lambda_b , \ \lambda_c]  \ \lambda_a)
\end{equation}
.and they are completely antisymmetric in any pair of indices. The
Gell-Mann
matrices also satify
\begin{equation}\label{ANTIC}
\{ \lambda_a, \ \lambda_b   \} = \frac{1}{3} \delta_{a b} {\bf {I}}
+  d_{a b
c}  \ \lambda_c,
\end{equation}
 where $d_{a b c}$ are completely symmetric in any pair of indices.

The adjoint representation of $SU(3)$ has dimension eight and
possesses two
invariant tensors
\begin{equation}\label{INVTEN}
\delta_{a b}= 2 tr (\lambda_a \lambda_b), \qquad d_{a b c}= 2 tr(
\{ \lambda_b,
\lambda_c \} \lambda_a).
\end{equation}

The irreducible representations of $SU(3)$ are labeled by two
integers $(p, q)$
where $p+q$
denotes the number of boxes in the first  row and $q$ denotes the
number of
boxes in the second row,  of the associated Young tableaux. The
dimension $D$
of the $(p,q)$ representation is
\begin{equation}\label{DIMREP}
D^{(p,q)}= \frac{1}{2}\left(p+q+2\right)\left(p+1\right)\left(q+1\right).
\end{equation}

The weights of the fundamental representation are
\begin{equation}\label{WEIGHTS}
\vec{\mu}_{1}=\left(\frac{1}{2},\frac{1}{2 \sqrt{3}} \right), \quad
\vec{\mu}_{2}=\left(-\frac{1}{2},\frac{1}{2 \sqrt{3}}\right), \quad
\vec{\mu}_{3}=\left(0,-\frac{1}{\sqrt{3}}\right),
\end{equation}
which denote de $\lambda_3$,  $\lambda_8$ eigenvalues respectively.

In the course of this research, sometimes it has proved useful to
rewrite the
partition function (\ref{WEYL1}) introduccing some standard polynomial
functions, like $F_1(k;u,v)$
\begin{equation}\label{F1}
u^{k+1}-v^{k+1}=(u-v)\sum_{j=0}^{k}u^j v^{k-j}:= (u-v) F_1(k;u,v),
\end{equation}
together with $F_2(k;u,v,w)$,
\begin{equation}\label{F2}
F_1(k;u,v)-F_1(k;u,w):=(v-w) F_2(k-1;u,v,w),
\end{equation}
which leads to the more explicit form
\begin{equation}\label{EXPLF2}
F_2(k;u,v,w)=\sum_{i=0}^{k} u^i F_1(k-i;v,w)=\sum_{i=0}^{k} u^{k-i}
F_1(i;v,w).
\end{equation}
In general, for higher values of the subindex $n=3,4,\dots$, the
corrresponding polynomials can be defined inductively by
\begin{equation}\label{Fn}
F_n(k; u, v, w, \dots)=\sum_{i=0}^{k}u^i F_{n-1}(k-i; v,w,\dots).
\end{equation}

Some of the properties of the F's are the following:

\noindent
(i) They are completely symmetric with respect to the exchange of any two
arguments.

\noindent
(ii) The difference between two F's which differ in just one argument
is given by
\begin{eqnarray}\label{PROPF}
&&F_n(k;u_1, \dots, u_n, u_{n+1})-
F_n(k;u_1, \dots, u_n, u_{n+2})\nonumber \\
&&=(u_{n+1}-u_{n+2})
F_{n+1}(k-1;u_1, \dots, u_n, u_{n+1},u_{n+2}).
\end{eqnarray}

In terms of the above polynomials $F_1$ and $F_2$, different cases
of the partition function (\ref{WEYL1}) are written as follows
\begin{equation}\label{Zpq0}
Z_0^{(p, q=0)}= F_2(p; \Lambda_1, \Lambda_2, \Lambda_3).
\end{equation}
\begin{equation}\label{Zp=q}
Z_0^{(p=q)}= F_1(p; \Lambda_1, \Lambda_2) F_1(p; \Lambda_2, \Lambda_3)
F_1(p; \Lambda_1, \Lambda_3)
\end{equation}
\begin{eqnarray}\label{ZGEN}
Z_0^{(p , q)}&=&\frac{1}{\Lambda_1-\Lambda_2}\left(
\Lambda_1^{q+1} F_1(q; \Lambda_2, \Lambda_3)
F_1(p; \Lambda_1, \Lambda_3)- \right. \nonumber \\
&& \left. \Lambda_2^{q+1} F_1(q; \Lambda_1, \Lambda_3)
F_1(p; \Lambda_2, \Lambda_3)\right)
\end{eqnarray}
\begin{equation}\label{ZGEN1}
Z_0^{(p,q)}=\sum_{j=0}^{q} \sum_{k=0}^{p}\Lambda_3^{j+k}
(\Lambda_1 \Lambda_2)^{q-j} F_1(p+j-k; \Lambda_1, \Lambda_2).
\end{equation}

Some explicit expressions of the F's polynomials are the
following
\begin{eqnarray}\label{EXF1}
&&F_1(1; u, v)= u + v, \quad F_1(2; u, v)= u^2 + uv + v^2, \nonumber \\
&&F_1(3; u, v)= u^3 + u^2 v + u v^2 + v^3,
\end{eqnarray}
\begin{eqnarray}\label{EXF2}
F_2(1; u, v, w)&=& u+v+w, \nonumber \\
F_2(2;u,v,w)&=& u^2+u^2+w^2 +uv + uw
+vw, \nonumber \\
F_2(3; u, v, w)&=& u^3 + v^3 +w^3 +u^2(v+w) + v^2(u+w) + w^2(u+v)
\nonumber \\
&+& uvw.
\end{eqnarray}

\vfill
\eject

%%%%%%%%%%%%%%%%%%%%%%%%%%%%%%%%%%%%%%%%%%%%%%
%%%%%%%%%%aqui termina zfin2.tex%%%%%%%%%%%%%%%%%%%%%%%%%
%%%%%%%%%%%%%%%%%%%%%%%%%%%%%%%%%%%%%%%%%%%%%%%

\newpage

{\LARGE Figure Captions }

\

{\bf Fig. 1 \ } General form of the specific heat as  
a function of temperature, for the different phases.

{\bf Fig. 2 \ } General form of  $\frac{\partial {\rm ln}
Z_0{}^{(p,q)}}{\partial
\sigma}$  as a function of $\sigma$, for $\lambda=0$.

{\bf Fig. 3 \ } Specific heat for  $ p=17, \;\; q=48 \longleftrightarrow
\mu= 40.0, \;\; \psi=45.0^{o}$. The upper curve is the classical result.
The temperature is in classical units.

{\bf Fig. 4 \ } Specific heat for  $ p=25, \;\; q=30 \longleftrightarrow
\mu=33.0, \;\; \psi=33.0^{o}$. The upper curve is the classical result.
The temperature is in classical units.

{\bf Fig. 5 \ } Specific heat for different values of  $ p=q
\longleftrightarrow \psi=30.0^{o}, \; {\rm different \ values \ of} \ \mu $. 
The upper curve is the classical result.
The temperature is in classical units.

{\bf Fig. 6 \ } Specific heat for $p=1 \;\; q=2
\longleftrightarrow \mu= 2,9 \;\; \psi=36.6^{o} $ and 
 $p=27,  \;\;
q=41 \longleftrightarrow \mu=40.7 \;\; \psi=36.6^{o}$. 
The upper curve is the classical result.
The temperature is in classical units.

%{\bf Fig. 7 \ } Classical entropy for  $\psi=43.0^{o} $.

{\bf Fig. 7 \ }  Quantum entropy for  $ p=25,  \;\;  q=30 $.

%{\bf Fig. 9 \ } Classical entropy for  $\psi=30^{o} $.

{\bf Fig.  8 \ } Quantum entropy for  $p=q $.
The temperature is in classical units.

%{\bf Fig. 11  \ }  Classical entropy for  $\psi=36.6^{o} $.

{\bf Fig. 9  \ } Quantum entropy for $\;\; p=1
\;\; q=2 $ and  $ p=27 \;\;  
q=41 \;\;$. The temperature is in classical units.

%{\bf Fig. 10 \ }  Classical order parameters for   

{\bf Fig. 10 \ } Quantum order parameters for  $ p=27,  
\;\; q=41
$.

{\bf Fig. 11 \ } Quantum order parameters for  $ p=1,  
\;\; q=2 $.

%%%%%%%%%%%%%%%%%%%%%%%%%%%%%%%%%%%%%%%%

\newpage

\

\vskip.5cm

\

\hskip 50pt \epsfxsize=3.5in \epsfbox{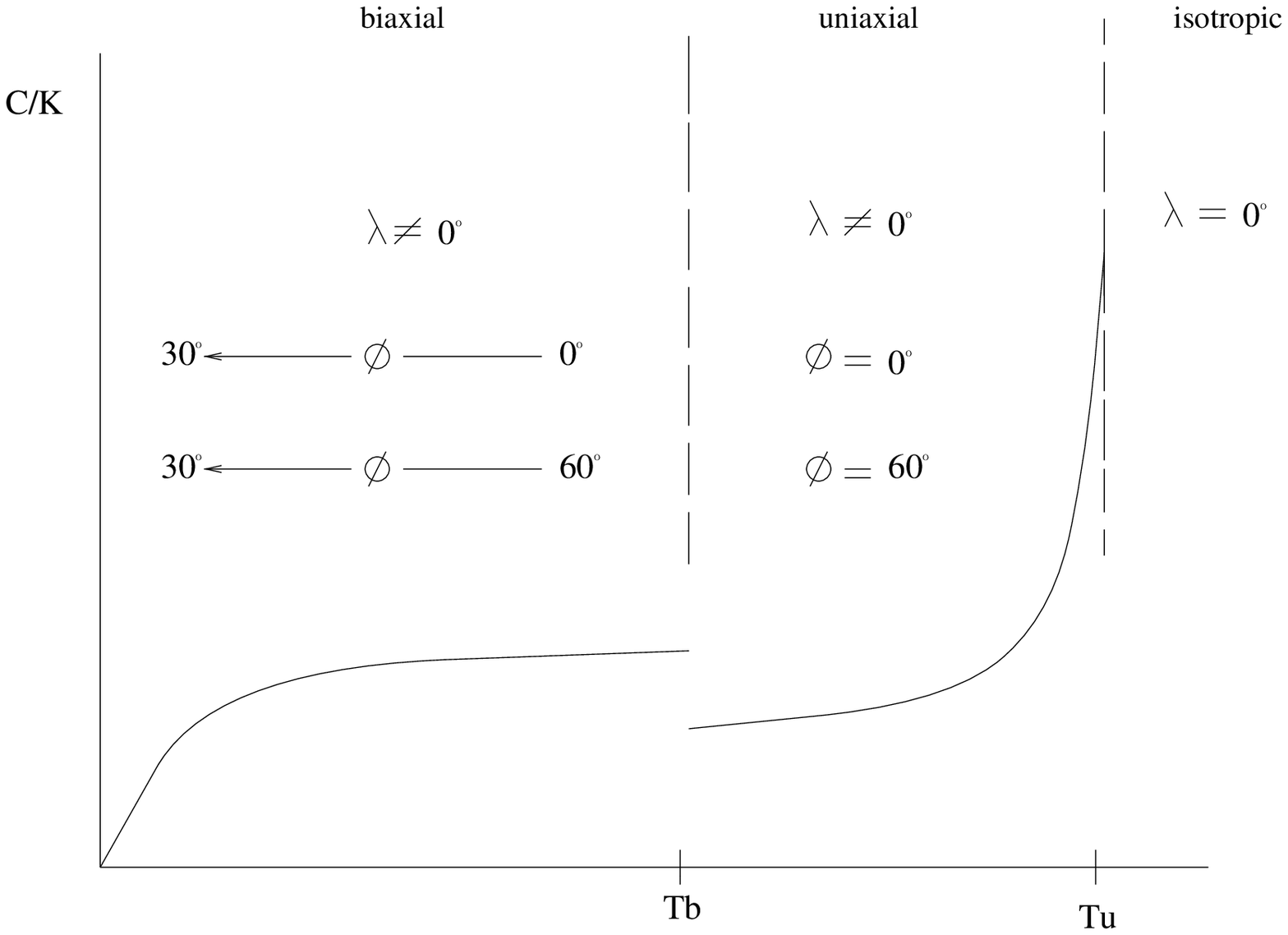}

\vskip.3cm

\centerline{  {\bf Fig. 1 \ }}

\

\vskip1.5cm

\hskip 50pt \epsfxsize=4in \epsfbox{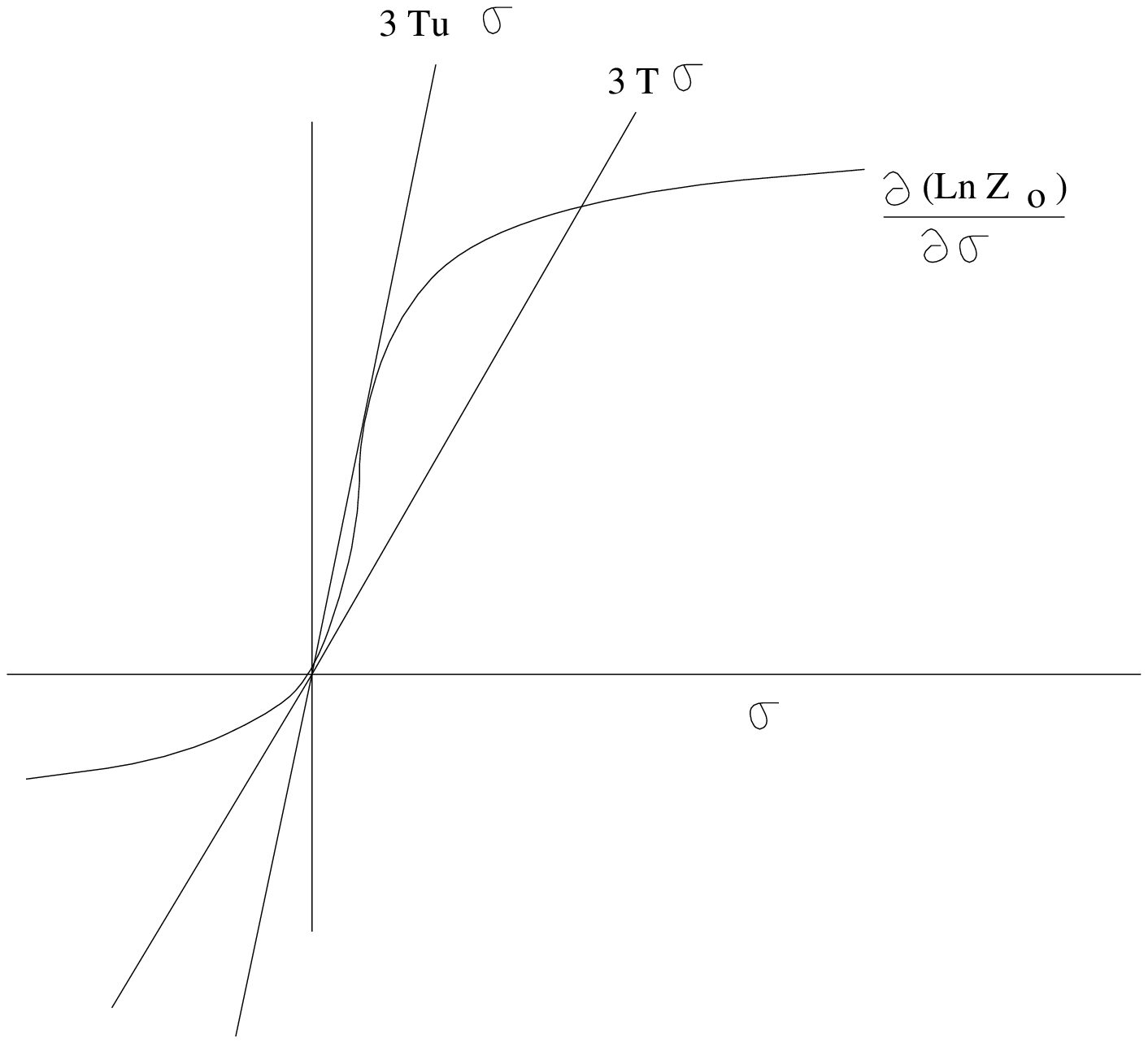}

\centerline{ {\bf Fig. 2 \ } }

\newpage

%{\LARGE Figures}

%\end{document}

\

\vskip .5cm

\epsfxsize=5in
\centerline{\epsfbox{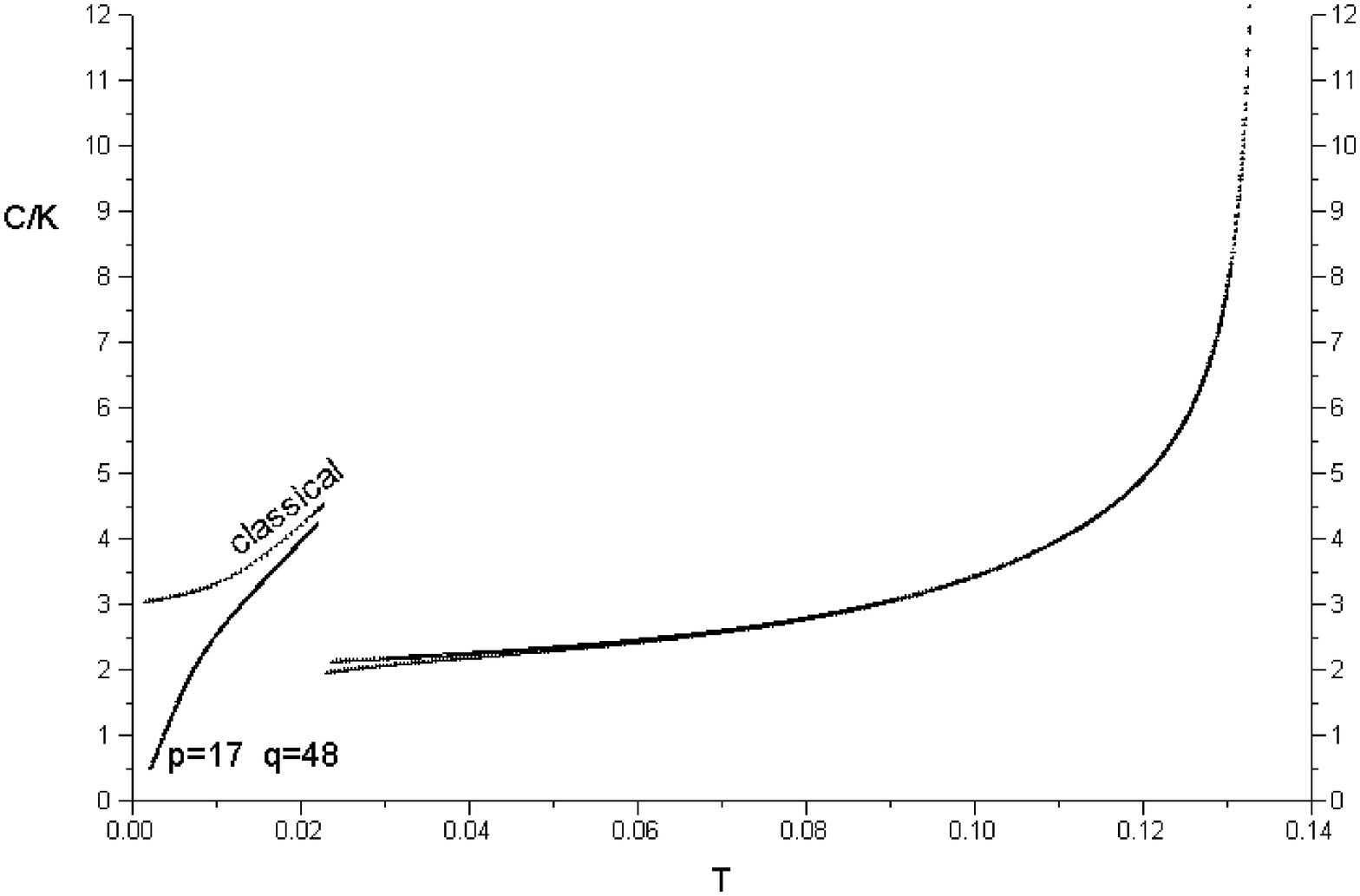}}
\centerline{{\bf Fig. 3 \ }}

\vskip1cm

\epsfxsize=5in
\centerline{\epsfbox{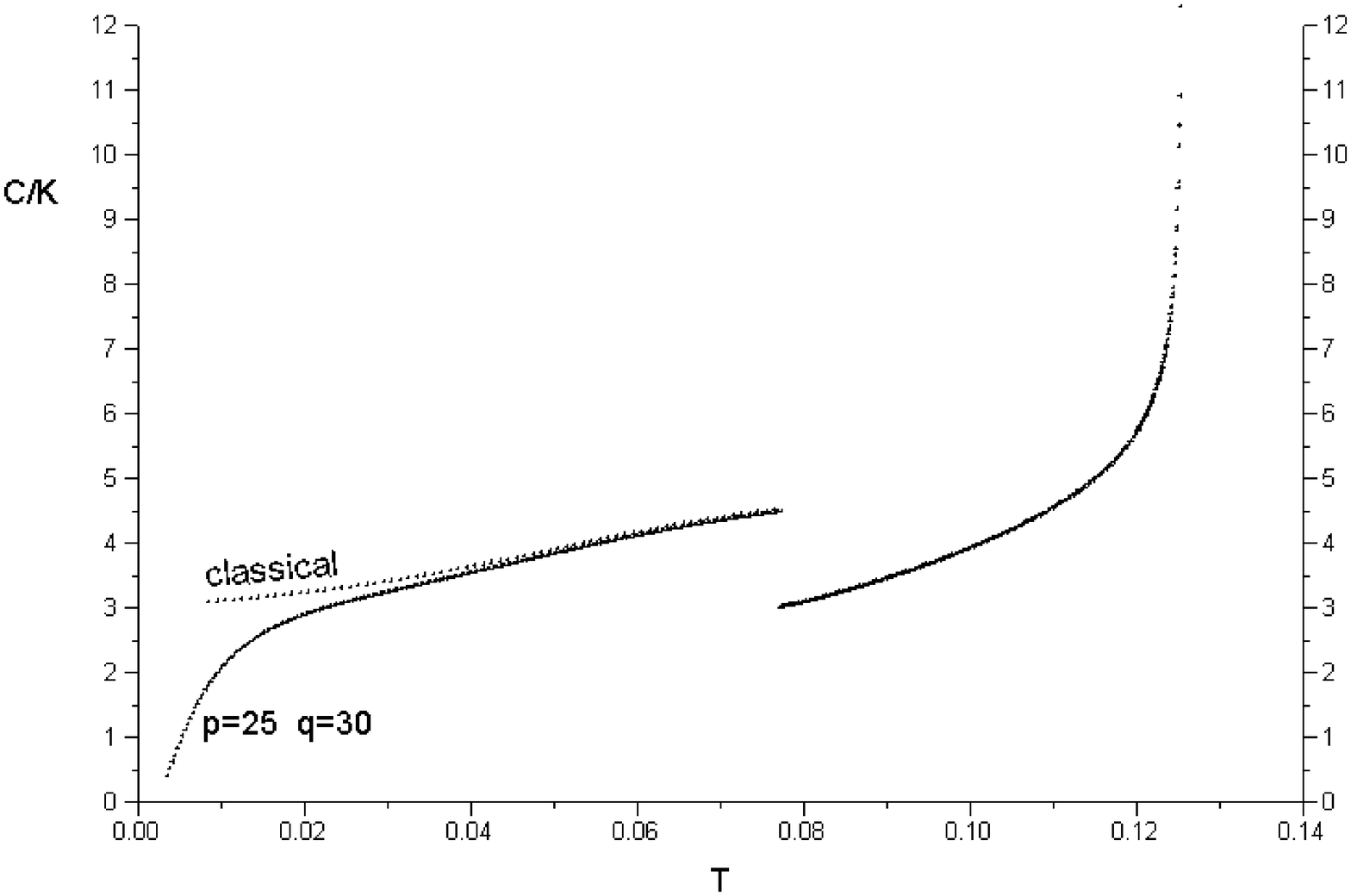}}
\centerline{ {\bf Fig. 4 \ }}

\newpage

\vskip.5cm

\epsfxsize=5in
\centerline{\epsfbox{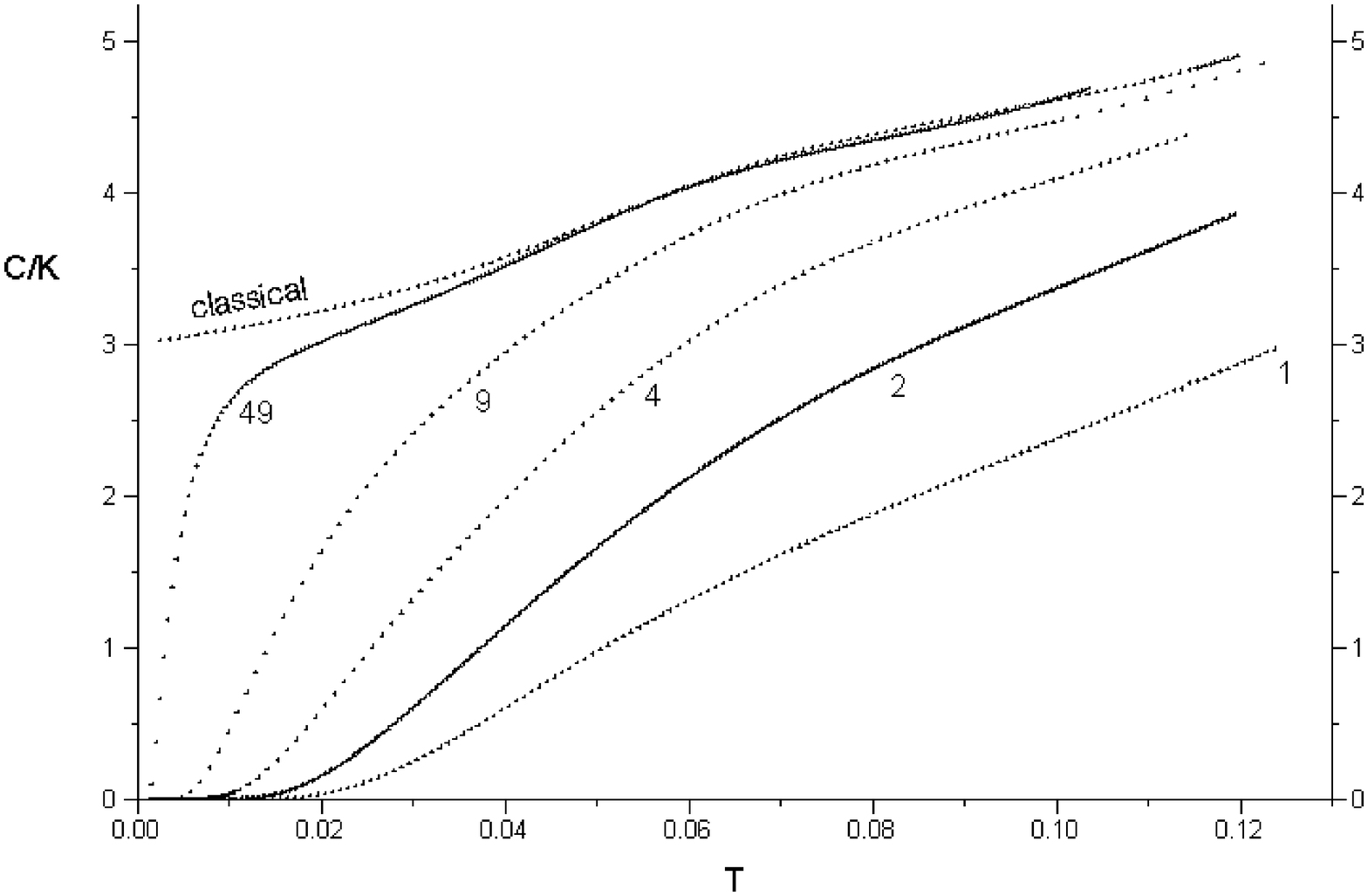}}
\centerline{{\bf Fig. 5 \ }}

\vskip1cm

\epsfxsize=5in
\centerline{\epsfbox{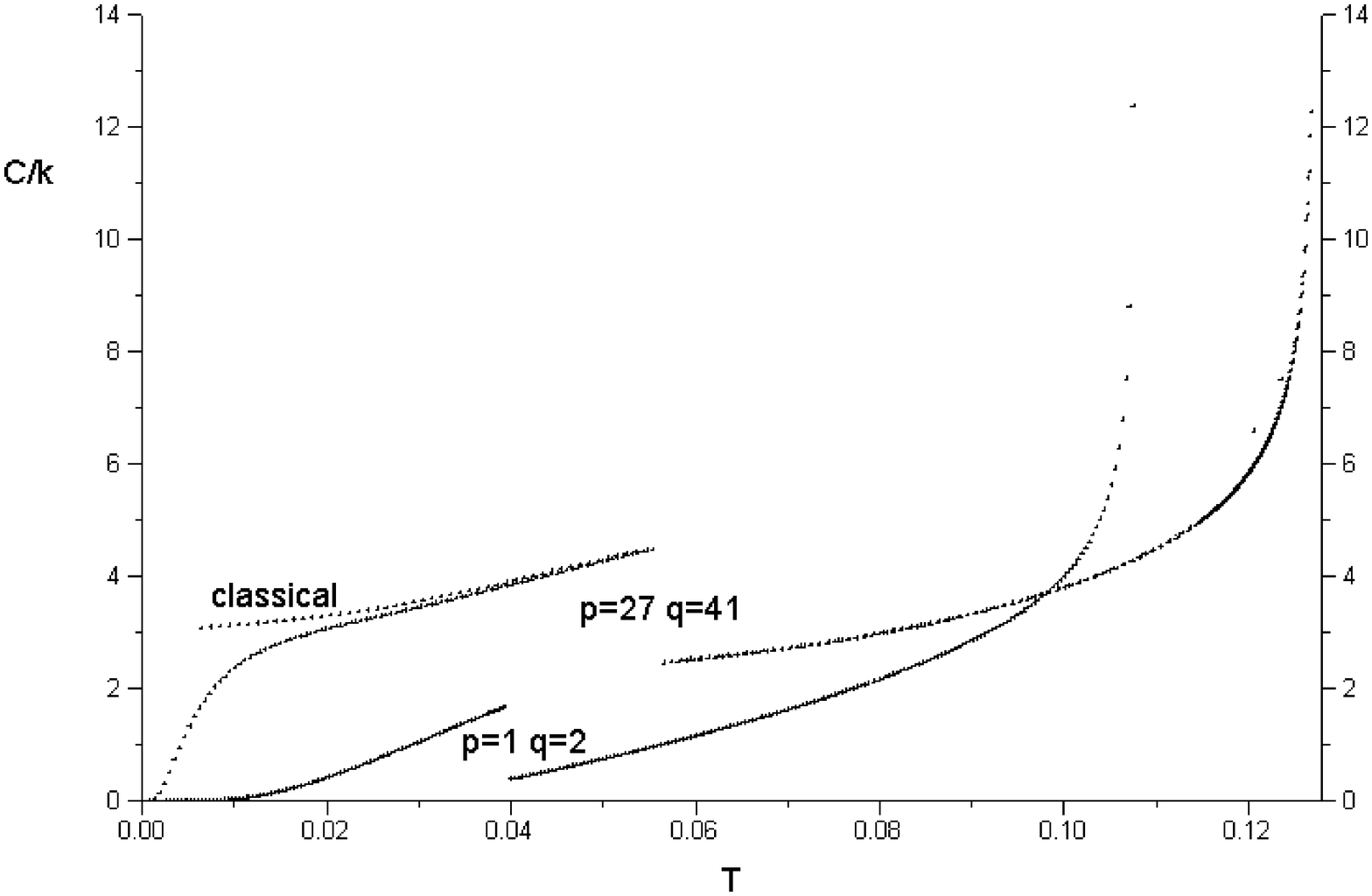}}
\centerline{ {\bf Fig. 6 \ }}

%\newpage

%\epsfxsize=6in
%\centerline{\epsfbox{sclas27.ps}}
%\centerline{ {\bf Fig. 7 \ }}

\newpage

\vskip.5cm

\epsfxsize=5in
\centerline{\epsfbox{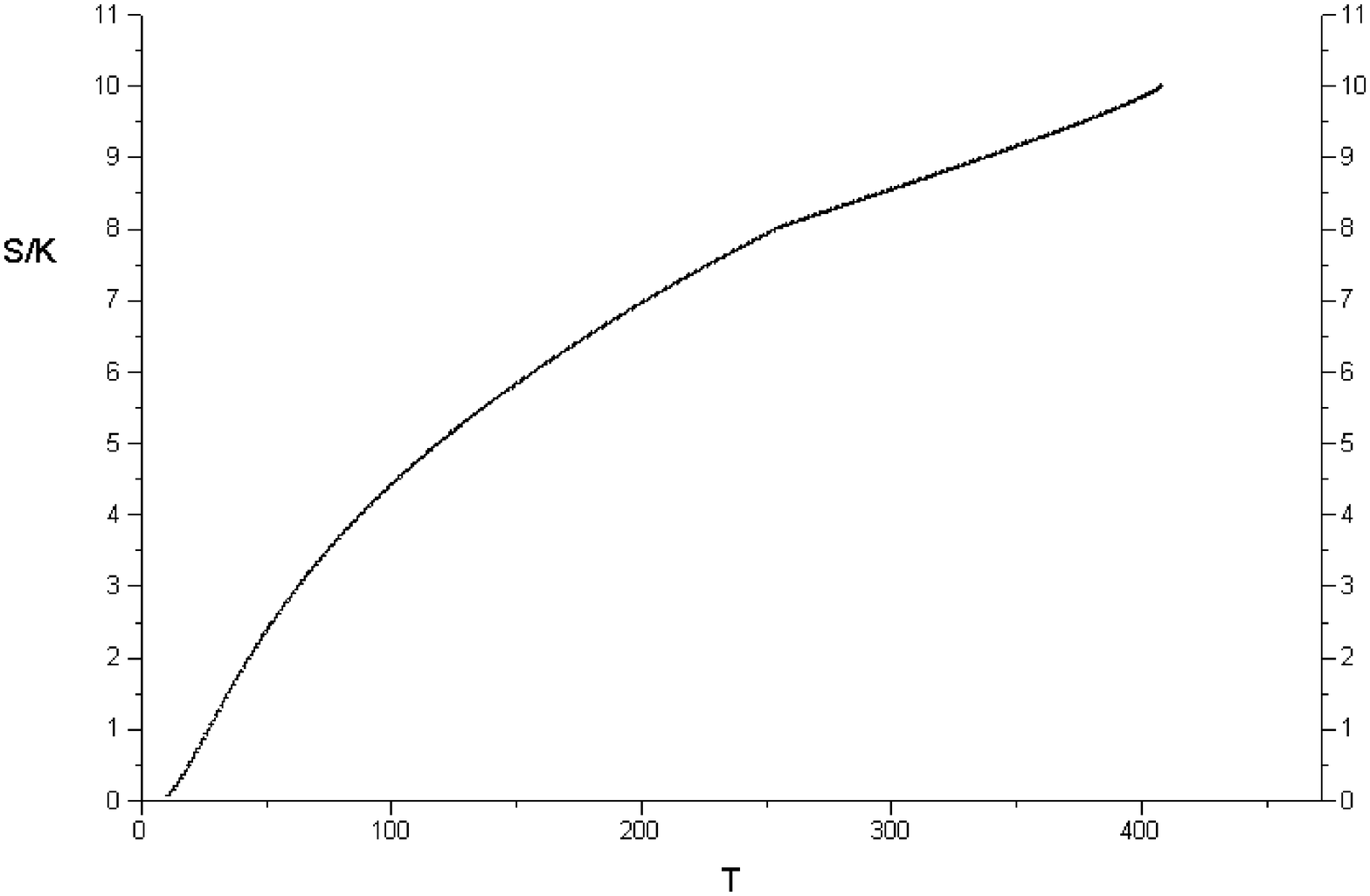}}
\centerline{{\bf Fig. 7 \ } }

%\newpage

%\epsfxsize=6in
%\centerline{\epsfbox{sclas30.ps}}
%\centerline{{\bf Fig. 9 \ }}

\vskip1cm

\epsfxsize=5in
\centerline{\epsfbox{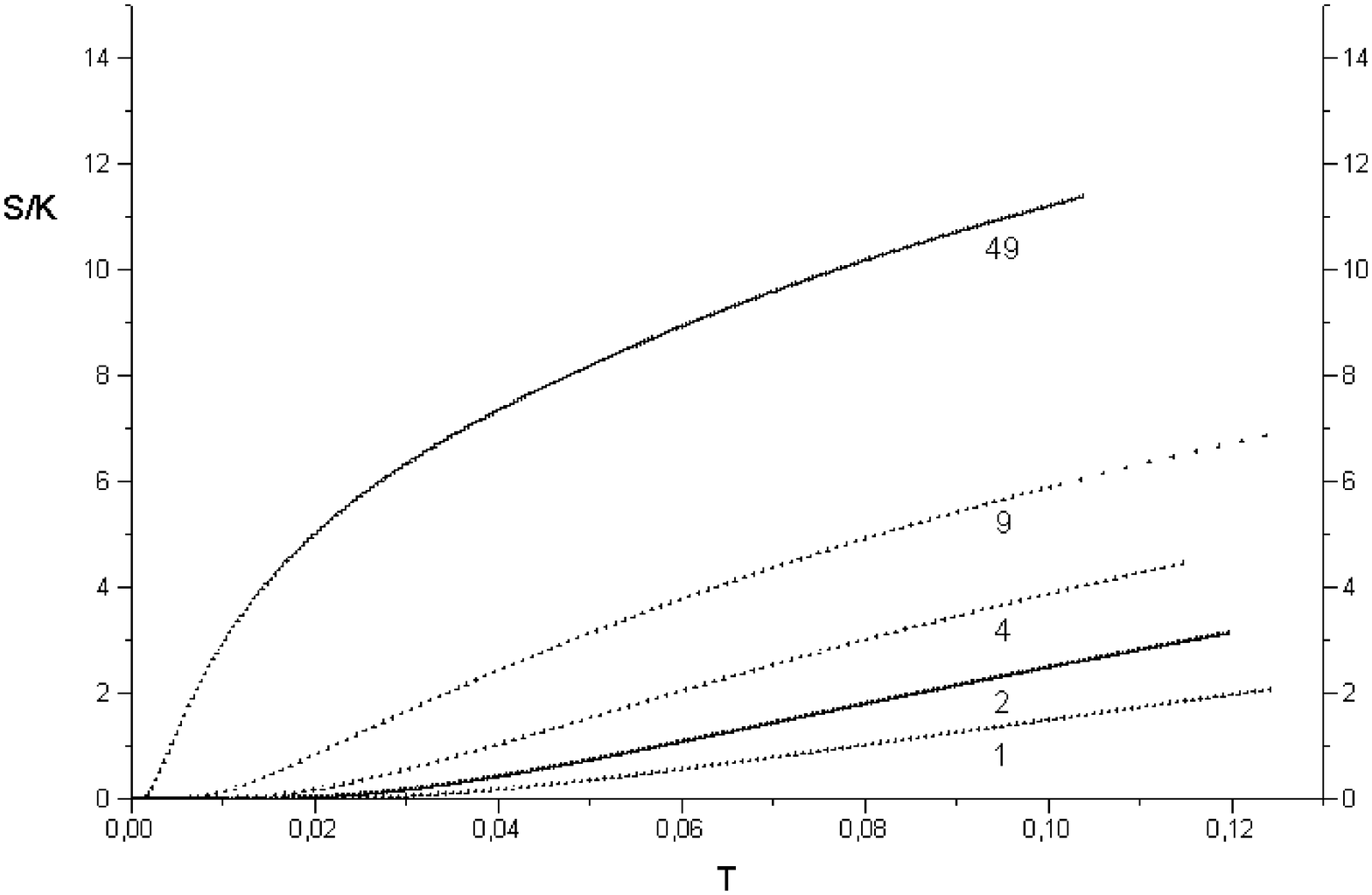}}
\centerline{{\bf Fig.  8 \ }}

%\newpage

%\epsfxsize=6in
%\centerline{\epsfbox{scla2341.ps}}
%\centerline{{\bf Fig. 11  \ }}

\newpage

\vskip.5cm

\epsfxsize=5in
\centerline{\epsfbox{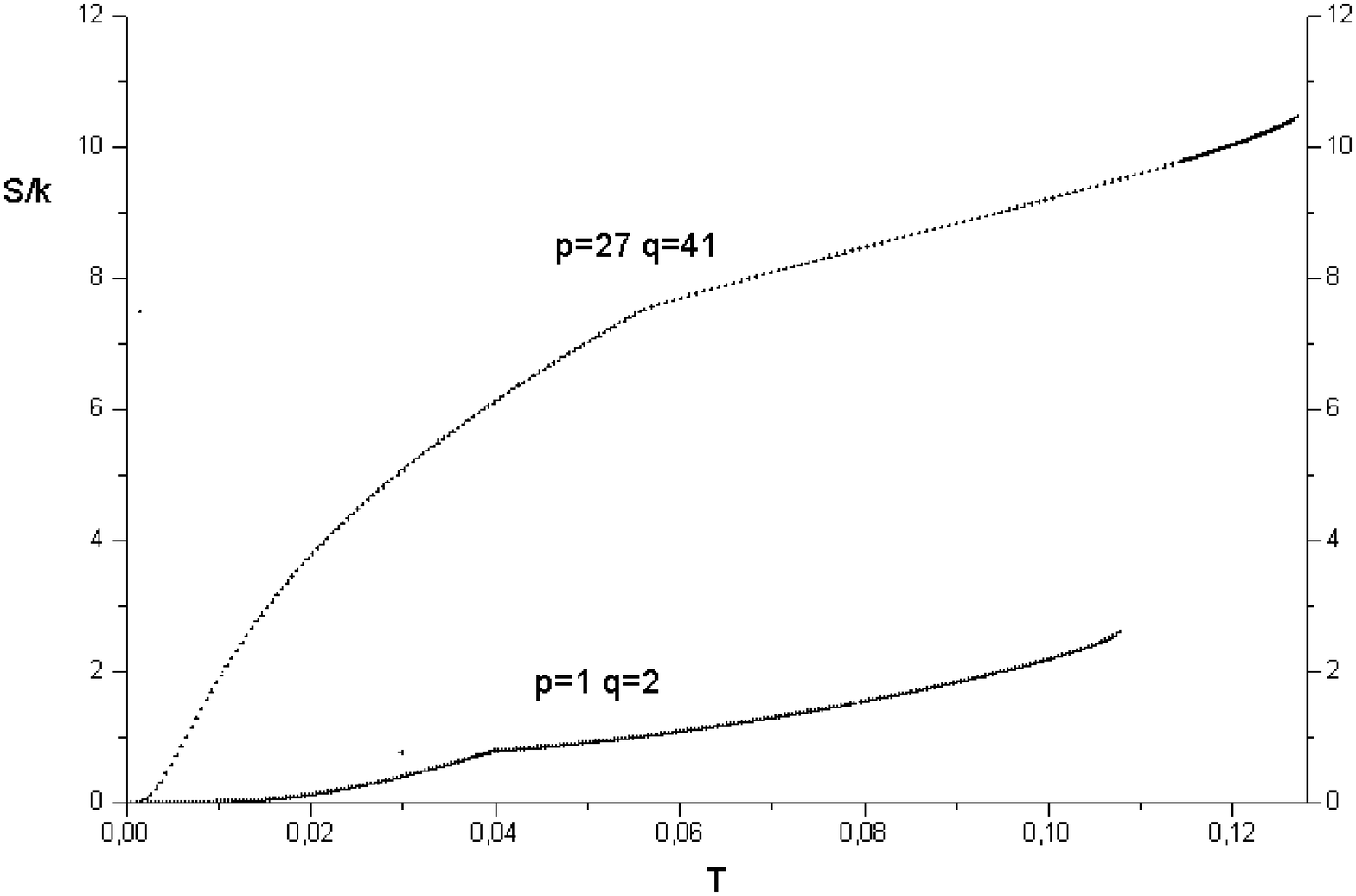}}
\centerline{ {\bf Fig. 9  \ }}

%\newpage

%\epsfxsize=6in
%\centerline{\epsfbox{pocl2341.ps}}
%\centerline{ {\bf Fig. 13 \ }}

\vskip1cm

\epsfxsize=5in
\centerline{\epsfbox{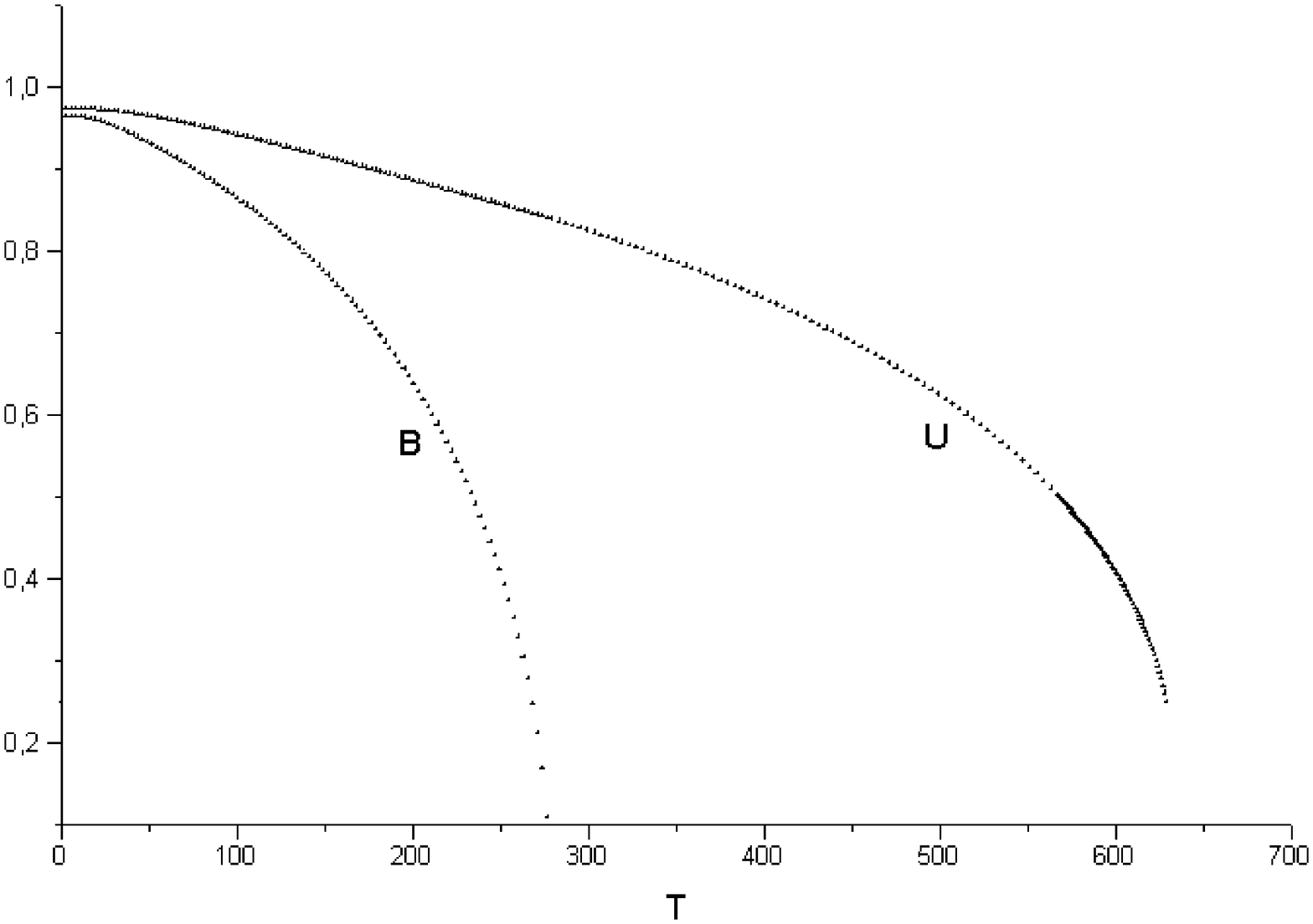}}
\centerline{ {\bf Fig. 10 \ }}

\newpage

\vskip.5cm

\epsfxsize=5in
\centerline{\epsfbox{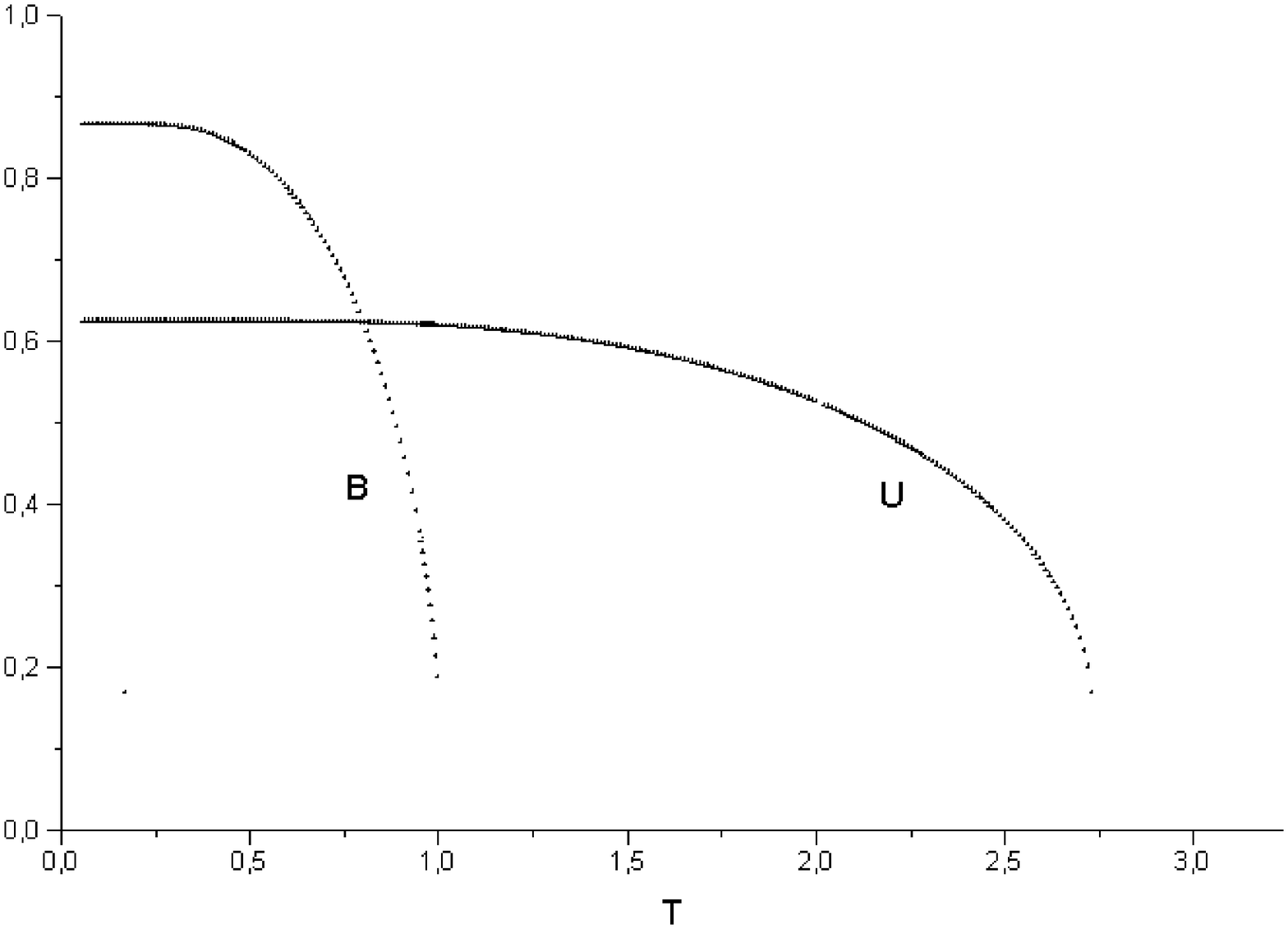}}
\centerline{ {\bf Fig. 11 \ }}

%%%%%%%%%%%%%%%%%%%%%%%%%%%%%%%%%%%%%%%%

\newpage

\

\vskip1.5cm

\centerline{\bf Table I  }

$$\begin{array}{llllllll}
p \;\;\;\;\;\;\; & q \;\;\;\;\;\;\; & \mu \;\;\;\;\;\;\;\;\; & \psi

\;\;\;\;\;\;\; & T_{b} & \rightarrow &

(T_{b})_c \;\;\;\;\;\;\; & T_{b}({\rm classical}) \\ \\
2 & 24 & \frac{2}{3}\sqrt{709} & 54.4 & 2.667 &   & 0.00282 &
0.00317 \\
4 & 43 & \frac{2}{3}\sqrt{2181} & 54.7 & 8.000 &   & 0.00275 &
0.00287 \\
17 & 48 & \frac{2}{3}\sqrt{3607} & 45.0 & 109.5 &   & 0.02277 &

0.02286
\\
1 & 2 & \frac{2}{3}\sqrt{19} & 36.6 & 1.015 &   & 0.04008 &
0.05614 \\
27 & 41 & \frac{28}{3}\sqrt{19} & 36.6 & 278.3 &   & 0.05605 &
0.05614 \\

27 & 47 & \frac{8}{3}\sqrt{277} & 38.6 & 273.5 &   & 0.04628 &
0.04632
\\
25 & 30 & \frac{2}{3}\sqrt{2443} & 32.9 & 256.1 &   & 0.07862 &
0.07877
\end{array}$$

\vskip2cm

\

\centerline{\bf Table II }

$$\begin{array}{lllllll}
p \;\;\;\;\;\;\; & q \;\;\;\;\;\;\; & \psi \;\;\;\;\;\;\;&
T_{u} & \rightarrow & (T_{u})_c \;\;\;\;

\;\;\; &
T_{u}({\rm classical}) \\ \\
2 & 24 & 54.4 & 129.5 &   & 0.1370 & 0.1454 \\
4 & 43 & 54.7 & 383.2 &   & 0.1318 & 0.1455 \\
17 & 48 & 45.0 & 637.8 &   & 0.1326 & 0.1361 \\
1 & 2 & 36.6 & 2.725 &   & 0.1076 & 0.1275 \\
27 & 41 & 36.6 & 631.8 &   & 0.1272 & 0.1275 \\
27 & 47 & 38.6 & 760.5 &   & 0.1287 & 0.1292 \\
25 & 30 & 32.9 & 408.2 &   & 0.1253 & 0.1255
\end{array} $$

\

\end{document}